\documentclass[acmtog,nonacm]{acmart}
\acmJournal{TOG}

\usepackage{amsmath}
\usepackage{graphics}

\usepackage{color}
\newcommand{\new}[1]{{\color{black} #1}}
\usepackage{wrapfig}
\usepackage{subfigure}
\usepackage[ruled]{algorithm2e}

\usepackage{amssymb}
\usepackage{placeins} 
\usepackage{makecell} 
\usepackage{booktabs}
\usepackage{multirow}
\usepackage{diagbox}

\usepackage{nicematrix}
\usepackage{caption}
\usepackage{enumitem}

\usepackage{listings}
\usepackage{caption}

\newcounter{nalg} 
\renewcommand{\thenalg}{A.\arabic{nalg}} 
\DeclareCaptionLabelFormat{algocaption}{Algorithm \thenalg} 

\lstnewenvironment{algorithm_mine_def}[1][] 
{   
    \refstepcounter{nalg} 
    \captionsetup{labelformat=algocaption,labelsep=colon} 
    \lstset{ 
        mathescape=true,
        frame=tB,
        numbers=left, 
        numberstyle=\tiny,
        basicstyle=\scriptsize, 
        keywordstyle=\color{black}\bfseries\em,
        keywords={,input, output, return, datatype, function, if, else, foreach, while, begin, end, } 
        numbers=left,
        xleftmargin=.04\textwidth,
        breaklines=true,
        #1 
    }
}
{}

\newtheorem{theorem}{Theorem}

\AtBeginDocument{%
  \providecommand\BibTeX{{%
    \normalfont B\kern-0.5em{\scshape i\kern-0.25em b}\kern-0.8em\TeX}}}

\begin{document}

\title[P2M: A Fast Solver for Querying Distance from Point to Mesh Surface]{P2M: A Fast Solver for Querying Distance from Point to Mesh Surface}


\author{Chen Zong}
\authornote{Both authors contributed equally to this research.}
\email{zongchen.official@qq.com}
\orcid{0000-0003-4954-0780}

\author{Jiacheng Xu}
\authornotemark[1]
\email{202000130031@mail.sdu.edu.cn}
\orcid{0009-0008-2855-7405}

\affiliation{%
  \institution{Shandong University}
  \city{Qingdao}
  \state{Shandong}
  \country{China}
}

\author{Jiantao Song}
\email{202120688@mail.sdu.edu.cn}
\orcid{0009-0007-0907-3731}
\affiliation{%
  \institution{Shandong University}
  \city{Qingdao}
  \state{Shandong}
  \country{China}
}

\author{Shuangmin Chen}
\email{csmqq@163.com}
\orcid{0000-0002-0835-3316}
\affiliation{%
  \institution{Qingdao University of Science and Technology}
  \city{Qingdao}
  \country{China}}

\author{Shiqing Xin}
\authornote{Corresponding author.}
\email{xinshiqing@sdu.edu.cn}
\orcid{0000-0001-8452-8723}
\affiliation{%
  \institution{Shandong University}
  \city{Qingdao}
  \state{Shandong}
  \country{China}
}

\author{Wenping Wang}
\email{wenping@tamu.edu}
\orcid{0000-0002-2284-3952}
\affiliation{%
  \institution{Texas A\&M University}
  \city{Texas}
  \country{USA}
}

\author{Changhe Tu}
\email{chtu@sdu.edu.cn}
\orcid{0000-0002-1231-3392}
\affiliation{%
  \institution{Shandong University}
  \city{Qingdao}
  \state{Shandong}
  \country{China}
}

\begin{abstract}

Most of the {existing} point-to-mesh distance query solvers, such as Proximity Query Package~(PQP), Embree and Fast Closest Point Query~(FCPW), are based on bounding volume hierarchy~(BVH). The hierarchical organizational structure enables one to eliminate the vast majority of triangles that do not help find the closest point.
In this paper, we develop a totally different algorithmic paradigm, named {\em P2M}, to speed up point-to-mesh distance queries. 
Our original intention is to precompute
a KD tree (KDT) of mesh vertices to approximately encode the geometry of a mesh surface containing vertices, edges and faces. 
However, 
it is very likely that the closest primitive to the query point is an edge~$e$ (resp., a face~$f$), 
but the KDT reports a mesh vertex~$v$ instead. 
We call~$v$ an {\em interceptor} of~$e$ (resp.,~$f$).
The main contribution of this paper is to invent a simple yet effective interception inspection rule and an efficient flooding interception inspection algorithm for quickly finding out all the interception pairs. 
Once the KDT and the interception table are precomputed,
the query stage proceeds by first searching the KDT and then looking up the interception table to retrieve the closest geometric primitive.
Statistics show that our query algorithm runs
many times faster than the state-of-the-art solvers. 
\end{abstract}




\maketitle


\section{Introduction}
Given a mesh surface,
fast query of the closest geometric primitive (vertex, edge, or face) to the user-specified point,
as well as the closest point and the minimum distance, 
is a fundamental operation in a wide range of research fields~\cite{wald2019rtx,guezlec2001meshsweeper,abbasifard2014survey,auer2013semi} 
including computer graphics, physical simulation, computational geometry and computer-aided design. 

Bounding volume hierarchy (BVH)~\cite{haverkort2004introduction} is a commonly used data structure to encode the hierarchical inter-primitive spatial proximity.
Once BVH is constructed in the preprocessing stage,
it greatly 
expedites the query
by quickly eliminating those triangles that do not help determine the minimum distance. 
 
\new{CGAL \cite{CGAL} includes an AABB-based point-to-mesh distance query function. 
A much faster BVH-based implementation~\cite{liu2010fast,wang2013thickening}
is based on the proximity query package (PQP)~\cite{Larsen1999FastPQ}, which uses oriented bounding boxes (OBBs) as the basic bounding volume type.} 
There are some accelerated versions such as FCPW~\cite{FCPW2021github} and Embree~\cite{afra2016embree}, 
but they do not really bring down the query cost. 

\begin{figure}[tb]
	\centering
	\includegraphics[width=0.85\columnwidth]{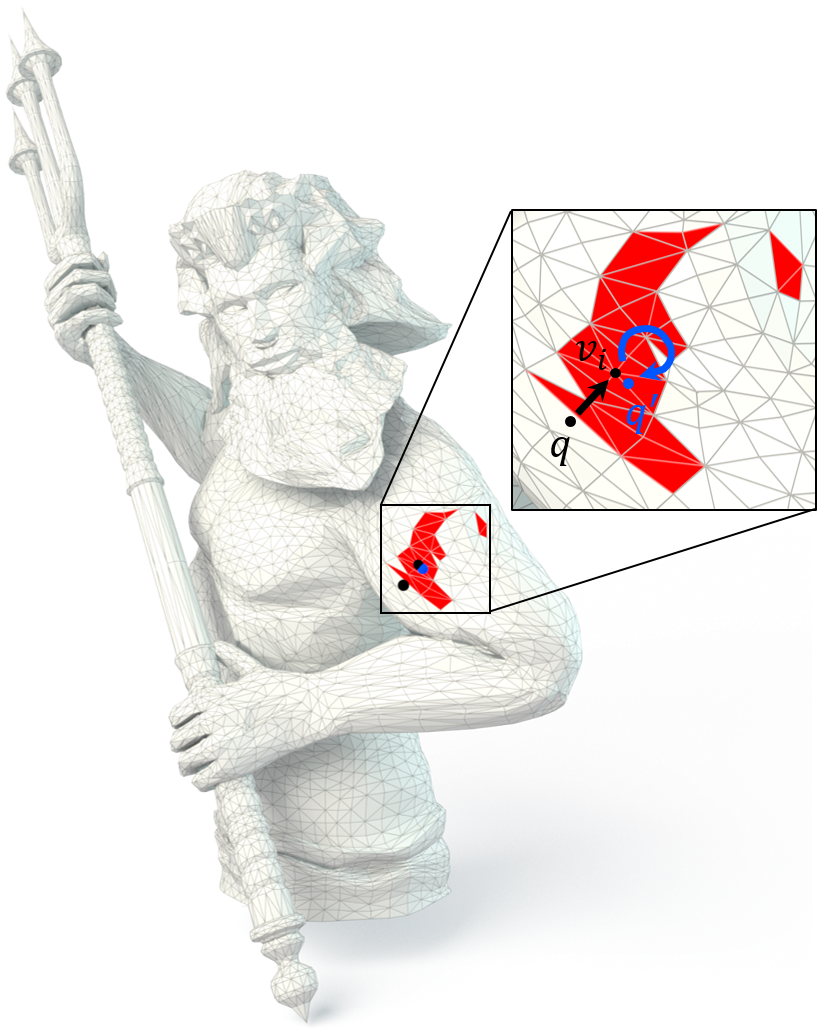}
	\caption{
	Given a query point~$q$, our query operation
	begins with searching the KDT of mesh vertices.
	Let $v_i$ be the nearest vertex to~$q$, reported by the KDT.
	After that, one needs to further look up $v_i$'s interception list and 
	finally identify the geometric primitive (an edge or a face)
	that contains the real closest point~$q'$.
	}
	\label{fig:intercept_poseidon}
	\vspace{-5mm}
\end{figure}

In this paper, we intend to use the KD tree (KDT), built from 
the vertex set~$V=\{v_i\}_{i=1}^n$, to 
help identify the real geometric primitive that defines the minimum distance.
When users input a query point~$q$ whose 
nearest point
is found to be~$v_i$ by KDT search,
we hope that the vertex~$v_i$ is able to keep enough clues to help find the real closest point~$q'$.
Suppose that the mesh edge~$e$ (resp., face~$f$)
is the geometric primitive containing~$q'$.
We say that~$v_i$ {\em intercepts}~$e$ (resp.,~$f$).
The main task of this paper is to precompute the interception table,
with which 
the query stage proceeds by first searching the KDT and then looking up the table to report the real closest point;
See Figure~\ref{fig:intercept_poseidon}.

Let $V,E,F$  be respectively the vertex set, the edge set and the face set of a triangle mesh.
Mathematically we take each vertex as a point with no dimension,
each edge as an open line segment with no width,
and each face as a bounded open region with no thickness. 
Just as the vertex set can define a Voronoi diagram~$\mathcal{V}_V$ in 3D, the geometric primitives in~$V,E,F$ can altogether define a generalized Voronoi diagram~$\mathcal{V}_{V,E,F}$, following the above definition.
Let $\textit{Cell}(v;\mathcal{V}_V)$ be $v$'s cell in~$\mathcal{V}_V$,
and $\textit{Cell}(e;\mathcal{V}_{V,E,F})$ be $e$'s cell
in~$\mathcal{V}_{V,E,F}$.
We observe that
$v$ {intercepts}~$e$ (resp.,~$f$)
if and only if the intersection between 
$\textit{Cell}(v;\mathcal{V}_V)$
and  $\textit{Cell}(e;\mathcal{V}_{V,E,F})$ (resp.,  $\textit{Cell}(f;\mathcal{V}_{V,E,F})$)
is not empty.

Based on this observation,
we give two techniques for fast interception inspection.
First, we relax the intersection domain (possibly non-convex) into a convex polytope

and give an effective filtering rule for fast interception inspection.
Second, we suggest a flooding procedure
of interception inspection to avoid exhausting all the vertex-edge and vertex-face pairs. 
The couple of techniques, simple yet effective,
enables one to precompute the interception table in a short period of time, e.g., about two minutes for a 1500K-face Dragon model. 
Note that the timing cost for accomplishing the same preprocessing task in a brute-force manner is more than one day!
We conduct extensive experiments to compare our algorithm 
with the BVH-based point-to-mesh distance query solvers. 
Experimental results show that our query,
with the support of the interception table,
is many times faster than the SOTA methods.

\section{Related Work}
Two topics are related to the theme of this paper, including nearest neighbor (NN) search and BVH.
\subsection{Nearest neighbor search}
Given a set of points in the $k$-dimensional space, 
NN search algorithms aim to find the one that is nearest to the user-specified input point~$q$. The search operation can be done efficiently by organizing the points into a tree such that large portions of the search space can be eliminated in the query stage. Most NN search algorithms include a tree construction stage and a tree-based query stage. 

\paragraph{KDT} 
Suppose that we have $N$ $k$-dimensional points~$P=\{p_i\}_{i=1}^N$, and each point of~$P$ has a form of $(x_1,x_2,\cdots,x_k)$. 
We first find the median point $p_i$ to divide the other $N-1$ points along the first dimension. 
The following process can be conducted in a divide-and-conquer fashion except that the division of points is done for different dimensions alternatively. The construction is finished when all the $N$ points are arranged in the tree.
In fact, each non-leaf node defines an axis-aligned hyperplane to split the space of interest into two parts.

In the query stage, the algorithm moves down the tree depending on the relative position of the query point to the splitting hyperplane. 
Once reaching a leaf node, the algorithm updates the best-so-far distance. 
Then it needs to unwind the recursion of the tree
and update the current best if there is another node that 
gives a smaller distance. 
Besides the task of querying the nearest point, 
the KDT can also be extended in several ways, e.g., searching $k$-nearest neighbors or retrieving points in a given hyperbox or hypersphere.
\paragraph{Other NN search data structures} In fact, there are many other data structures~\cite{li2019approximate} devised for NN search, for example, R-tree~\cite{guttman1984r,beckmann1990r,kamel1993hilbert,berchtold1996x}, ball-tree~\cite{liu2006new}, A-tree~\cite{sakurai2000tree}, BD-tree~\cite{white1996similarity}, SR-tree~\cite{katayama1997sr}
and Voronoi diagrams.
Although Voronoi diagrams encode the proximity between points more precisely, 
 the construction/search of Voronoi diagram is not easy~\cite{VoRTree2010,DEVROYE200461}
 especially with the increase of point dimensions.
 Furthermore, the operation of locating a point
 in a Voronoi diagram is less efficient than the KDT.

\subsection{Bounding volume hierarchy}
Like the KDT, a BVH~\cite{haverkort2004introduction} is a hierarchical structure to encode a set of geometric primitives. Each leaf node wraps a single geometric primitive while each non-leaf node keeps the enclosing bounding volume of a subset of geometric primitives. 
The BVH can be built in a top-down, bottom-up, or incremental insertion-based style, ultimately producing 
a tree  with a bounding volume at the top. 
BVH has been widely used in distance query~\cite{ytterlid2015bvh}, ray tracing~\cite{meister2021survey} and collision detection~\cite{funfzig2006hierarchical,wang2021review}.  

\paragraph{Bounding volume types} In the past research, various types of bounding volumes have been proposed and tested. 
The choice of bounding volume is a trade-off between
simplicity and tightness. On one hand, a simple bounding volume enables fast intersection tests and distance computation. 
On the other hand, the bounding volume is expected to fit the enclosed geometric primitives as tight as possible. Commonly used bounding volumes include AABB~\cite{beckmann1990r,larsson2006dynamic},
bounding sphere\\~\cite{palmer1995collision,hubbard1995collision,kavan2005fast}, OBB~\cite{gottschalk1996obbtree} and K-DOP~\cite{klosowski1998efficient}.
More bounding structures include zonotope~\cite{guibas2003zonotopes}, pie slice~\cite{barequet1996boxtree}, ellipsoid~\cite{liu2007ellipsoid}, VADOP~\cite{coming2007velocity} and convex hull~\cite{ehmann2001accurate}.

\paragraph{PQP} 
Considering that the typical discrete representation of a 3D object
is a triangle mesh or a triangle soup, 
the task of encoding the spatial proximity is to arrange a collection of triangles into a BVH.
The PQP~\new{\cite{Larsen1999FastPQ},
as well as the modified version~\cite{liu2010fast,wang2013thickening},} exploits
OBBs to wrap geometric primitives,
facilitating fast distance query. 
Statistics show that PQP filters out the vast majority of triangles that do not help determine the minimum distance.

\paragraph{BVH acceleration techniques}
Besides some works that focus on improving the quality of bounding volumes,
there are some accelerated versions such as FCPW~\cite{FCPW2021github} and Embree~\cite{afra2016embree}. 
For example, FCPW applies a wide BVH with vectorized traversal to accelerate its queries to geometric primitives~\cite{FCPW2021github}.
While both SIMD~\cite{ruipu2010study} and
GPGPU~\cite{tang2011collision} have been used to improve the performance,
the required computational amount is not really reduced. 

\section{Insight}\label{sec:Insight}
In this section, we provide the insight based on the 2D setting,
but it is worth noting that the proposed algorithm in this paper is mainly devised for the 3D situation. 
\begin{figure}[t]
    \centering

    \includegraphics[width=\columnwidth]{"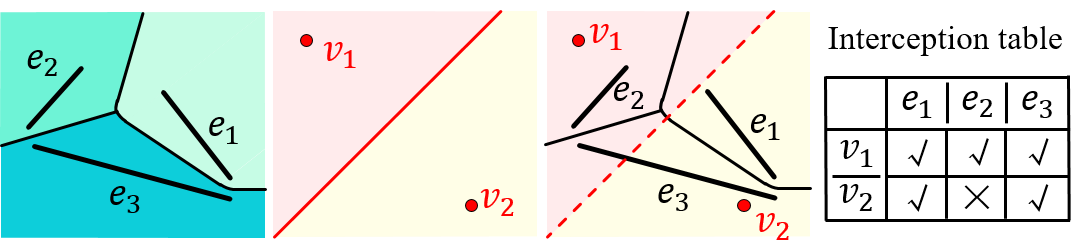"}
    \makebox[0.25\columnwidth][c]{(a)}\makebox[0.25\columnwidth][c]{(b)}\makebox[0.25\columnwidth][c]{(c)}\makebox[0.25\columnwidth][c]{(d)}
    \caption{
    Insight about the interception table.
    (a)~The Voronoi diagram w.r.t.~three segments $e_1,e_2,e_3$.
    (b)~The Voronoi diagram w.r.t.~two points $v_1, v_2$.
    (c)~Overlay visualisation of the two Voronoi diagrams.
    (d)~Interception table.
    Intuitively speaking, the interception table
encodes the principal-agent relationship between
a set of points and a set of more complicated geometric primitives.
    }
    \label{fig:insight}
    \vspace{-3mm}
\end{figure}

If the given geometric objects consist of finitely many discrete 2D points~$V=\{v_i\}_{i=1}^n$, there are some tree-type structures (like the KDT) 
to organize them to facilitate fast query of the nearest point. 
In fact, the point set $V$ determines a 2D Voronoi diagram~$\mathcal{V}_V$ that partitions the 2D plane into 
$n$
cells. The task of finding the nearest point to the query point~$q$ with the help of a KDT is equivalent to the task of locating~$q$ in $\mathcal{V}_V$. 

In our scenario, however, the geometric objects may be
more complicated. 
As Figure~\ref{fig:insight}(a) shows,
we have three line segments $E=\{e_1,e_2,e_3\}$. 
We need to report which line segment can provide the smallest
distance for a query point $q$ in the 2D plane. 
Similarly, $E$ also induces a Voronoi diagram~$\mathcal{V}_E$ where different Voronoi cells are visualized in different colors. 
The task of querying the closest point to \new{$q$ in} $E$,
in its nature,
is to locate the query point in~$\mathcal{V}_E$.

\new{
The Voronoi diagram w.r.t.~non-point geometric primitives
has curved bisectors, which are non-trivial to compute. 
Furthermore, even if the generalized Voronoi diagram has been computed, 
it is hard for one to quickly locate the query point. 
This motivates us to convert the point-to-mesh distance query problem to the traditional nearest point search problem where
the geometric primitives consist of finitely many points. 
} 
As Figure~\ref{fig:insight}(b) shows, 
we sample two points~$V=\{v_1,v_2\}$ in the 2D plane.
The Voronoi diagram~$\mathcal{V}_V$ is simply a straight-line bisector. 
Note that $V$ is not directly tied to~$E$ in this example
although~$V$ can be chosen to be a subset of~$E$ in practice.

The key idea of this paper
is to speed up the process of finding the closest geometric primitive in~$E$ with the help of a KDT of~$V$.

Figure~\ref{fig:insight}(c) gives an overlay visualization 
of~$\mathcal{V}_V$ and~$\mathcal{V}_E$.
Suppose that $v_1$ is nearer to $q$ than $v_2$. 
Then $q$ must be in the $v_1$'s cell of~$\mathcal{V}_V$.
Under this circumstance, the closest line segment to $q$ may be $e_1$ or $e_2$ or $e_3$.
If $v_2$ is nearer to $q$ than $v_1$, instead,
the closest line segment may be $e_1$ or $e_3$.
To summarize, $v$ is said to {\em intercept} an edge-type primitive~$e$
if and only if the following search space is non-empty:
\begin{equation}\label{eq:def}
\textit{Cell}(v;\mathcal{V}_V)\cap
\textit{Cell}(e;\mathcal{V}_{E})\neq\emptyset.
\end{equation}
We can also say that $v$ is an  {\em interceptor} of $e$
if the interception occurs.
In this way, we obtain an {\em interception} table to keep the interception relationship between $V$ and $E$; See Figure~\ref{fig:insight}(d). 

Roughly speaking, 
the KDT encodes how the mesh vertices are positioned in the 3D space while the interception table
encodes the principal-agent relationship between
a set of points and a set of more complicated geometric primitives. 

\begin{figure}[ht]
    \centering
    \includegraphics[width=0.40\columnwidth]{"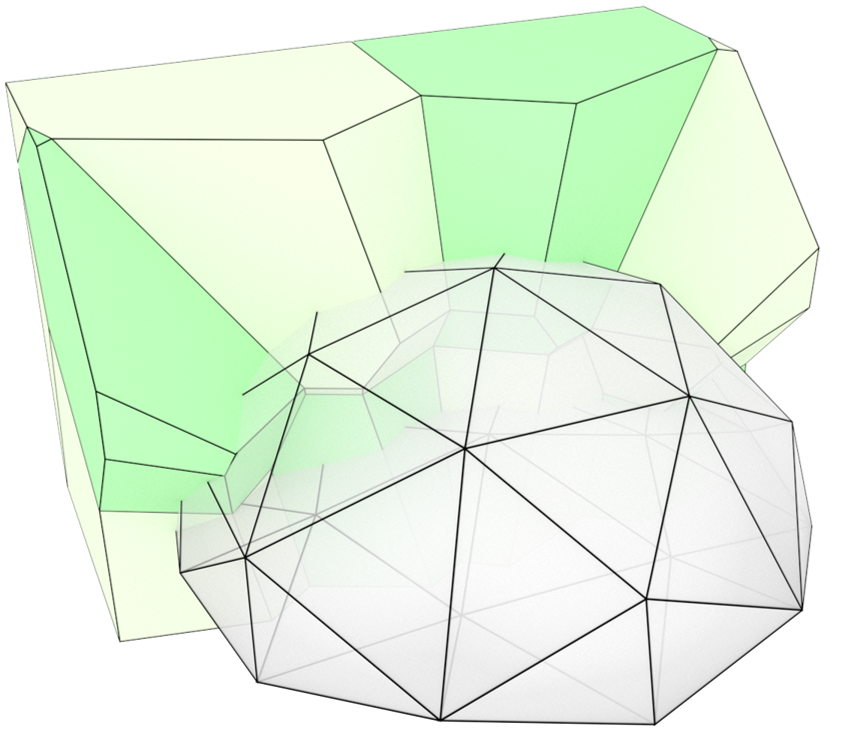"}
    \hspace{2mm}
    \includegraphics[width=0.40\columnwidth]{"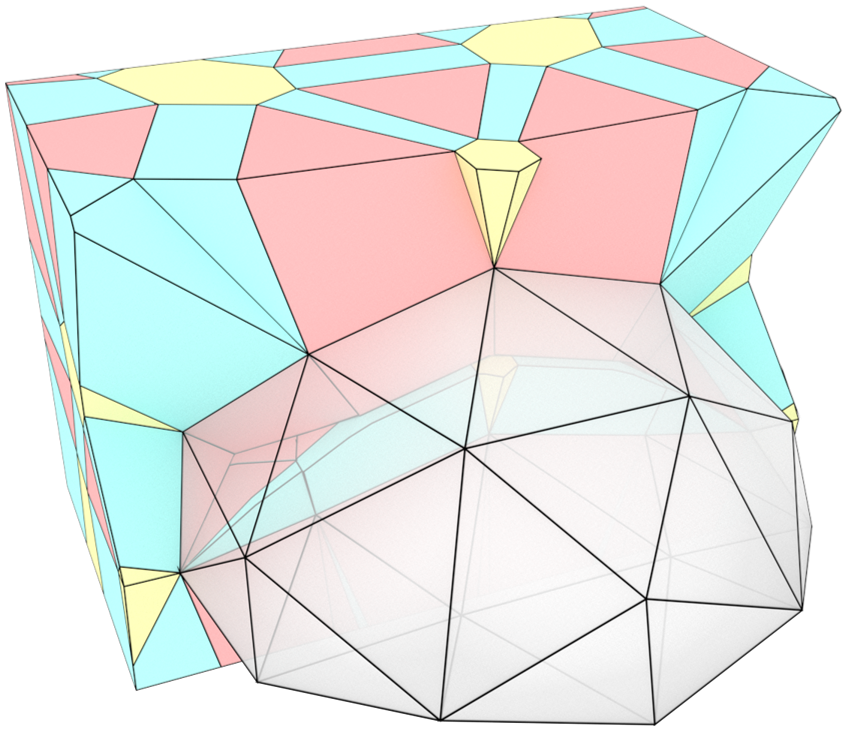"}
    \makebox[0.40\columnwidth][c]{(a)}
    \hspace{2mm}
    \makebox[0.40\columnwidth][c]{(b)}
    \caption{
    Given a mesh surface, the geometric primitives include the vertex set~$V$, the edge set~$E$ and the face set~$F$. 
    By taking each edge and face to be open (the endpoints or the boundary excluded),
    we precompute an interception table to
     keep the clues of how $V$ is tied to $E$ and $F$.
    (a)~The Voronoi diagram induced by~$V$, where only points are generators.
    (b)~The Voronoi diagram induced by~$V,E,F$, where points, edges and faces are all generators.
    }
    \label{fig:two_voronoi}
    \vspace{-3mm}
\end{figure}

\section{Formulation} \label{sec:Algorithm}

\subsection{Problem statement}
Suppose that we have a collection of triangles~$F=\{f_i\}_{i=1}^m$.
Rather than simply take $F$ as a triangle soup, 
we assume that $F$ owns only one copy for each vertex.
Let $V$ and $E$ be respectively the vertex set and the edge set.
We take the vertex set $V$ as the generators to define the Voronoi diagram~$\mathcal{V}_{V}$
while taking $V,E,F$ as the generators simultaneously to define the Voronoi diagram~$\mathcal{V}_{V,E,F}$,
where each edge/face is taken as an open point set (the endpoints or the boundary excluded);
See Figure~\ref{fig:two_voronoi} for illustration.
Based on the observation in Section~\ref{sec:Insight}, 
the difference between~$\mathcal{V}_{V}$ and~$\mathcal{V}_{V,E,F}$
induces the interception table,
i.e.,
a vertex $v$ intercepts an edge-type primitive $e$
if and only if the following search space is non-empty:
\begin{equation}
\label{eq:intersection:v:e}
\begin{aligned}
\textit{Cell}(v;\mathcal{V}_V)\cap
\textit{Cell}(e;\mathcal{V}_{V,E,F})\neq\emptyset.
\end{aligned}
\end{equation}
Similarly, 
$v$ intercepts a face-type primitive $f$
if and only if the intersection is non-empty:
\begin{equation}
\label{eq:intersection:v:f}
\begin{aligned}
\textit{Cell}(v;\mathcal{V}_V)\cap
\textit{Cell}(f;\mathcal{V}_{V,E,F})\neq\emptyset.
\end{aligned}
\end{equation}

However, 
$\textit{Cell}(e;\mathcal{V}_{V,E,F})$ and $\textit{Cell}(f;\mathcal{V}_{V,E,F})$ may have a 
curved boundary surface,
making it non-trivial 
to determine whether the intersection domain is empty,
which motivates us to study the structural features of $\textit{Cell}(e;\mathcal{V}_{V,E,F})$ and $\textit{Cell}(f;\mathcal{V}_{V,E,F})$,
and develop a fast interception inspection algorithm. 

\subsection{Structure of $\textit{Cell}(e;\mathcal{V}_{V,E,F})$ and $\textit{Cell}(f;\mathcal{V}_{V,E,F})$} \label{sec:intercept_3d}

\begin{figure}[htb]
    \centering
    \includegraphics[width=0.65\columnwidth]{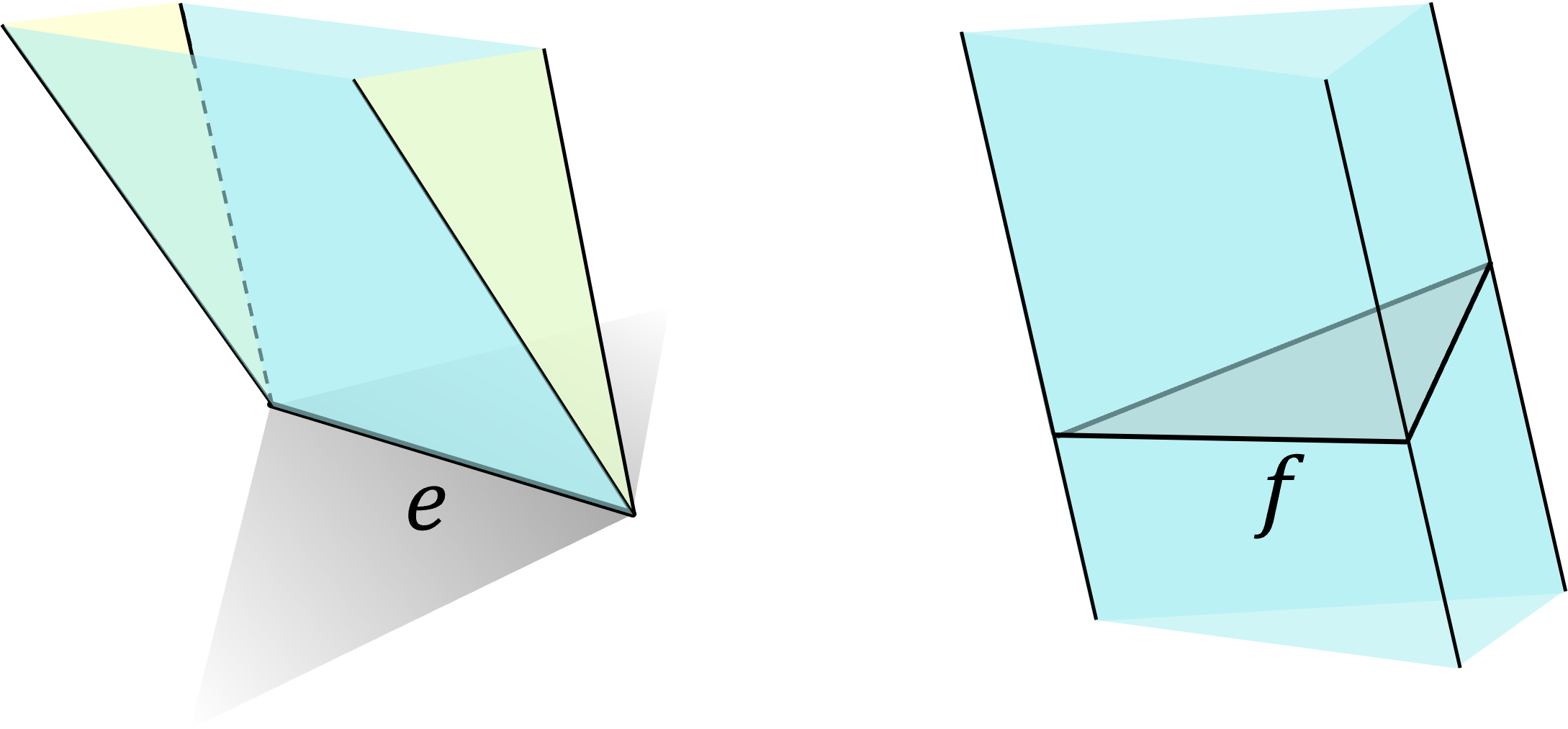}
    \makebox[0.36\columnwidth][c]{(a)}\makebox[0.36\columnwidth][c]{(b)}
    \caption{Vertical space for an edge (a) and a face (b) respectively.}
    \label{fig:perp_region}
    \vspace{-3mm}
\end{figure}

Suppose that we have a triangle face~$f\in F$. 
Then $\textit{Cell}(f;\mathcal{V}_{V,E,F})$ 
contains all the points that are nearer to $f$ than to any other geometric primitive. 
Recall that $f$ does not include its boundary edges.
If we project a point $q\in \textit{Cell}(f;\mathcal{V}_{V,E,F})$
onto the plane of~$f$,
then the projection $q'$ must be an interior point of~$f$, satisfying~$qq'\perp f$.
Therefore, we use the three bounding edges of~$f$
to define three vertical planes, and denote the space sandwiched  
by the three vertical planes by~$\textit{Space}^\perp(f)$,
as Figure~\ref{fig:perp_region}(b) shows. 
Obviously, we have
\begin{equation}\label{eq:f:perp}
    \textit{Cell}(f;\mathcal{V}_{V,E,F})\subset \textit{Space}^\perp(f).
\end{equation}
For an edge~$e$, $\textit{Space}^\perp(e)$ can be defined similarly.
As Figure~\ref{fig:perp_region}(a) shows,
the edge~$e$ is adjacent to two faces,
each of which defines a half space. 
Also, there are two half planes 
rooted at the endpoints of~$e$.
The intersection domain 
by the four half spaces 
defines $\textit{Space}^\perp(e)$.
Note that if~$e$ is adjacent to more or less than two faces, $\textit{Space}^\perp(e)$ can also be well defined. 

Likewise, we have
\begin{equation}\label{eq:e:perp}
    \textit{Cell}(e;\mathcal{V}_{V,E,F})\subset \textit{Space}^\perp(e).
\end{equation}

By combining $\textit{Cell}(v_j;\mathcal{V}_{V,E,F})
\subset \textit{Cell}(v_j;\mathcal{V}_{V})$
and $\textit{Cell}(v_i;\mathcal{V}_{V})\cap \textit{Cell}(v_j;\mathcal{V}_{V})=\emptyset$, 
we have the following observation. 
\begin{theorem}\label{thm:interception:impossible}
Suppose that $v_i,v_j\in V, i\neq j$. Then we have
$$\textit{Cell}(v_i;\mathcal{V}_{V})\cap \textit{Cell}(v_j;\mathcal{V}_{V,E,F})=\emptyset.$$
\end{theorem}

Furthermore, 
as the cell of $\textit{Cell}(v_i;\mathcal{V}_{V})$  
encloses the vertex $v_i\in V$,
the intersection between $\textit{Cell}(v_i;\mathcal{V}_{V})$ 
and every incident edge/face cannot be empty,
which shows that 
$v_i$ must intercept the incident edges and faces.

\begin{theorem}\label{thm:interception:must}
Suppose that $v_i\in V$ is a vertex. Then
$v_i$ must intercept the edges and the faces incident to~$v_i$. 
\end{theorem}

It's worth noting that 
$\textit{Space}^\perp(e)$
and
$\textit{Space}^\perp(f)$
are convex polyhedral domains, 
which can be represented by a collection of linear constraints. 
\subsection{Interception filtering} \label{subsec:interceptionfilter}
\label{subsec:Optimization}

Suppose that $v\in V, e\in E$, and we come to discuss in what situation~$v$ intercepts~$e$. 
Let~$l_e$ be the straight line of~$e$.
If~$v$ is an endpoint of~$e$, then~$v$ must be an interceptor of~$e$.

\begin{figure}[htb]
    \centering
    \includegraphics[width=0.88\columnwidth]{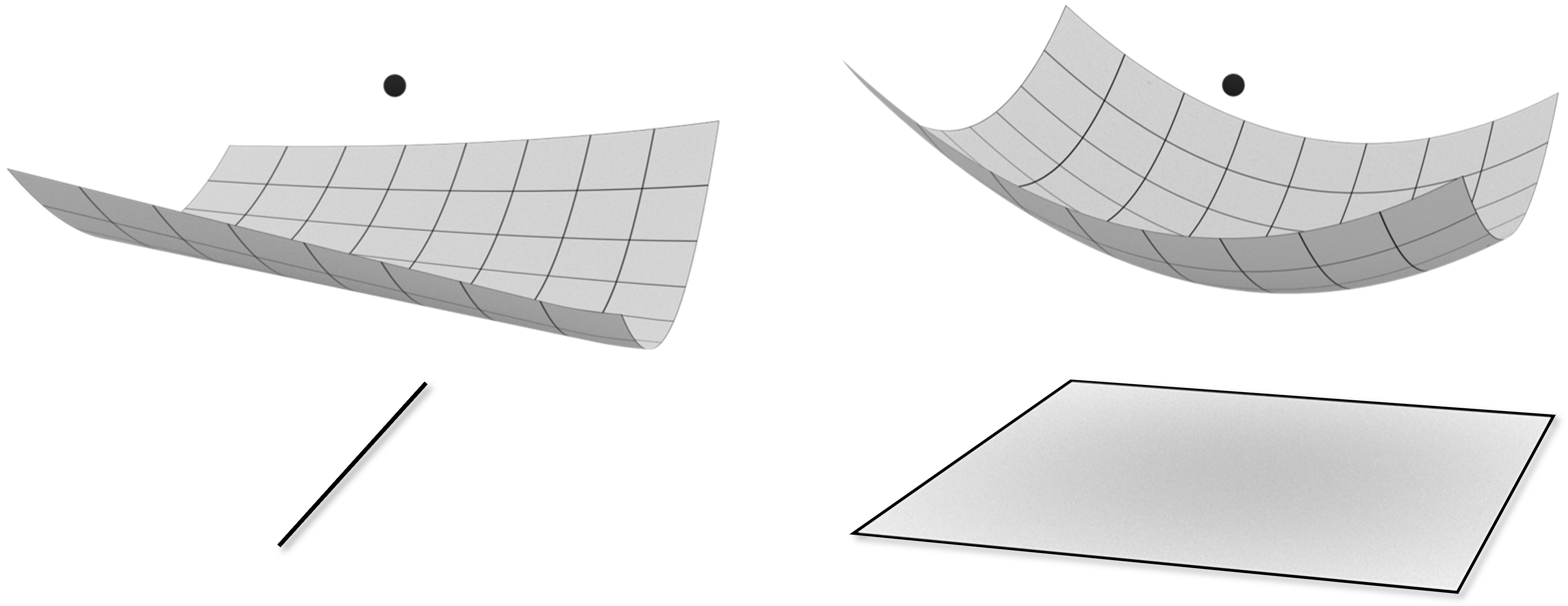}
    \makebox[0.44\columnwidth][c]{(a)}\makebox[0.44\columnwidth][c]{(b)}
    \caption{
    (a)~The bisector surface between a point and a straight line.
    (b)~The bisector surface between a point and a plane.
    For both cases, the bisector surface divides the whole space into a convex part and a non-convex part, where the point lies in the convex part.}
    \label{fig:bisector}
    \vspace{-3mm}
\end{figure}
\begin{figure}[htb]
    \centering
     \includegraphics[width=0.45\columnwidth]{"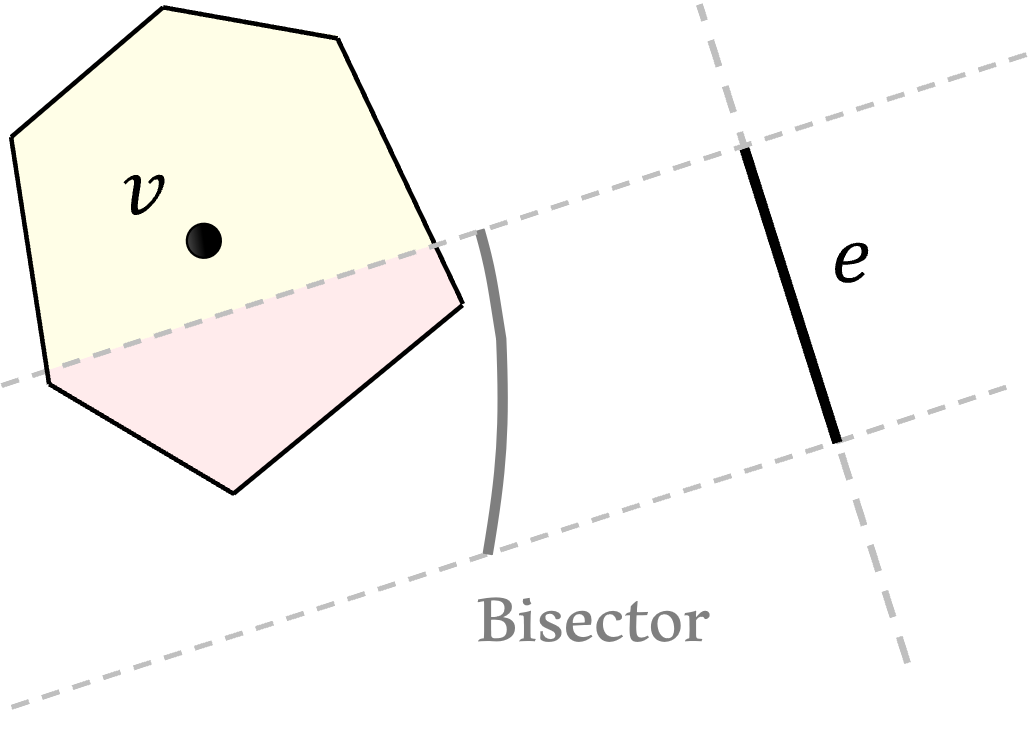"}
    \caption{Theorem~\ref{thm:Bisect:e} asserts that if the intersection domain~$\textit{Cell}(v;\mathcal{V}_{V})\cap \textit{Space}^\perp(e)$, colored in pink, belongs to $\textit{Bisect}^v(v,l_e)$, then $v$ 
cannot intercept~$e$.}
    \label{fig:non_intercept}
    \vspace{-3mm}
\end{figure}

As Figure~\ref{fig:bisector}(a) shows,
the point~$v$ and the straight line~$l_e$ determine a bisector surface 
that divides the whole space into two parts, where the part containing $v$ is convex. 
We use $\textit{Bisect}^v(v,l_e)$ to denote the convex part containing~$v$
while using $\textit{Bisect}^e(v,l_e)$ to denote the other part. 
The following theorem gives a situation that~$v$ cannot intercept~$e$; See Figure~\ref{fig:non_intercept}
for 2D illustration.

\begin{theorem}\label{thm:Bisect:e}
If $\textit{Cell}(v;\mathcal{V}_{V})\cap \textit{Space}^\perp(e)\subset\textit{Bisect}^v(v,l_e)$,
then $v$ 
cannot intercept~$e$.
\end{theorem}
\begin{proof}
If $\textit{Cell}(v;\mathcal{V}_{V})$ does not intersect the vertical space of~$e$, i.e., $$\textit{Cell}(v;\mathcal{V}_{V})\cap \textit{Space}^\perp(e)=\emptyset,$$ 
then we have $$\textit{Cell}(v;\mathcal{V}_{V})\cap \textit{Cell}(e;\mathcal{V}_{V,E,F})=\emptyset$$
due to~$$\textit{Cell}(e;\mathcal{V}_{V,E,F})\subset\textit{Space}^\perp(e).$$
Otherwise, any point $$q\in\textit{Cell}(v;\mathcal{V}_{V})\cap \textit{Space}^\perp(e)$$
must belong to $q\in\textit{Bisect}^v(v,l_e)$, which implies that $q$ is closer to~$v$ than to~$l_e$.

Under the assumption that~$v$ intercepts~$e$, there is at least one point
$$q'\in\textit{Cell}(v;\mathcal{V}_{V})\cap\textit{Cell}(e;\mathcal{V}_{V,E,F})\subset \textit{Cell}(v;\mathcal{V}_{V})\cap \textit{Space}^\perp(e)$$
such that the projection~$q''$ of $q'$ onto $l_e$ is located between the two endpoints of~$e$ (see Eq.~(\ref{eq:e:perp})). Furthermore, we have~$q'q''\perp l_e$ and $\|q'q''\|<\|q'v\|$, which contradicts the fact that any point in $\textit{Bisect}^v(v,l_e)$
is closer to~$v$ than to~$l_e$. 
\end{proof}

Next, we come to discuss in what situation a vertex~$v$ intercepts a triangle~$f$. 
We suppose that
the plane of~$f$ is~$\pi_f$.
The point~$v$ and the plane $\pi_f$ determine a bisector surface, as is shown in Figure~\ref{fig:bisector}(b). 
It is easy to filter out the following interception case
in a similar inference procedure.
\begin{theorem}\label{thm:Bisect:f}
If $\textit{Cell}(v;\mathcal{V}_{V})\cap \textit{Space}^\perp(f)\subset\textit{Bisect}^v(v,\pi_f)$,
then $v$ 
cannot intercept~$f$.
\end{theorem}

{\noindent\bf Remark. }
$\textit{Cell}(v;\mathcal{V}_{V})$ is a convex polygonal space but may be unbounded. 
In practical occasions, we can assume that both the vertex set and the query point are in a limited range, e.g., 
\begin{equation}
[-M, M]\times [-M, M]\times[-M, M],
\end{equation}
where $M$ is a sufficiently large constant.
We can add 8 virtual points into~$V$, yielding~$\overline{V}$:
\begin{equation}
\overline{V}=V\cup(\pm 3M, \pm 3M, \pm 3M).
\end{equation}
It can be proved that the nearest point to any query point $q$ in the limited range 
can only be a point in~$V$, 
 but not a virtual point.
The benefit of augmenting $V$ to $\overline{V}$
is that in the Voronoi diagram~$\mathcal{V}_{\overline{V}}$,
the cell of each $v\in V$ becomes bounded. 
Therefore, in what follows, we take $\textit{Cell}(v;\mathcal{V}_{V})$, as well as $\textit{Cell}(v;\mathcal{V}_{V})\cap \textit{Space}^\perp(e)$
or
$\textit{Cell}(v;\mathcal{V}_{V})\cap \textit{Space}^\perp(f)$, as a bounded convex polytope. 
\subsection{Convexity based filtering rule} \label{subsec:convexityfilter}
Theorem~\ref{thm:Bisect:e}
points out a situation of impossible interception, i.e.,
if $\textit{Cell}(v;\mathcal{V}_{V})\cap \textit{Space}^\perp(e)\subset\textit{Bisect}^v(v,l_e)$,
then $v$ 
cannot intercept~$e$,
where $\textit{ConvexPoly}(v,e)\triangleq\textit{Cell}(v;\mathcal{V}_{V})\cap \textit{Space}^\perp(e)$ defines a convex and bounded polytope.
Suppose that~the polytope of~$\textit{ConvexPoly}(v,e)$ has $k$ extreme points $\{x_i\}_{i=1}^k$. 
Due to the convexity of $\textit{ConvexPoly}(v,e)$ and $\textit{Bisect}^v(v,l_e)$,
the assertion of $\textit{ConvexPoly}(v,e)\subset\textit{Bisect}^v(v,l_e)$ is equivalent to
\begin{equation}
x_i\in \textit{Bisect}^v(v,l_e),\quad \forall i=1,2,\cdots,k.
\end{equation}
Based on the fact, we propose a convexity based filtering rule
as follows.
\begin{theorem}\label{thm:convexity:e}
We assume that $\textit{ConvexPoly}(v,e)$ has $k$ extreme points $\{x_i\}_{i=1}^k$ to define the convex volume. 
If 
\begin{equation}
\|x_i-v\|\leq \text{Dist}(x_i,l_e),\quad \forall i=1,2,\cdots,k,
\end{equation}
then $v$ 
cannot intercept~$e$,
where $\text{Dist}(x_i,l_e)$ denotes the distance between the point $x_i$ and the straight line~$l_e$.
\end{theorem}
\begin{theorem}\label{thm:convexity:f}
We assume that $\textit{ConvexPoly}(v,f)$ has $k$ extreme points $\{x_i\}_{i=1}^k$ to define the convex volume. 
If 
\begin{equation}
\|x_i-v\|\leq \text{Dist}(x_i,\pi_f),\quad \forall i=1,2,\cdots,k,
\end{equation}
then $v$ 
cannot intercept~$f$,
where $\text{Dist}(x_i,\pi_f)$ denotes the distance between the point $x_i$ and the plane~$\pi_f$.
\end{theorem}

\paragraph{Implementation}
To this end,
it is necessary to compute the convex polytope $\textit{ConvexPoly}(v,e)$ or $\textit{ConvexPoly}(v,f)$
to facilitate interception inspection.
In fact, the operation of generating a convex polytope by plane cutting is available in CGAL~\cite{CGAL} or Geogram~\cite{Geogram}, 
but our scenario is specific since the number of half-planes is quite limited. 

Therefore, we implement plane cutting by ourselves for consideration of run-time performance.

$\textit{ConvexPoly}(v,e)$
(or $\textit{ConvexPoly}(v,f)$) is initialized to be
$v$'s Voronoi cell, i.e.,
$\textit{Cell}(v;\mathcal{V}_V)$.
During the construction of $\textit{ConvexPoly}(v,e)$,
the convex polytope is repeatedly cut by
a sequence of half-planes. 
\begin{wrapfigure}{r}{0.45\columnwidth}
\vspace{-10pt}
\hspace{-20pt}
\includegraphics[width=0.50\columnwidth]{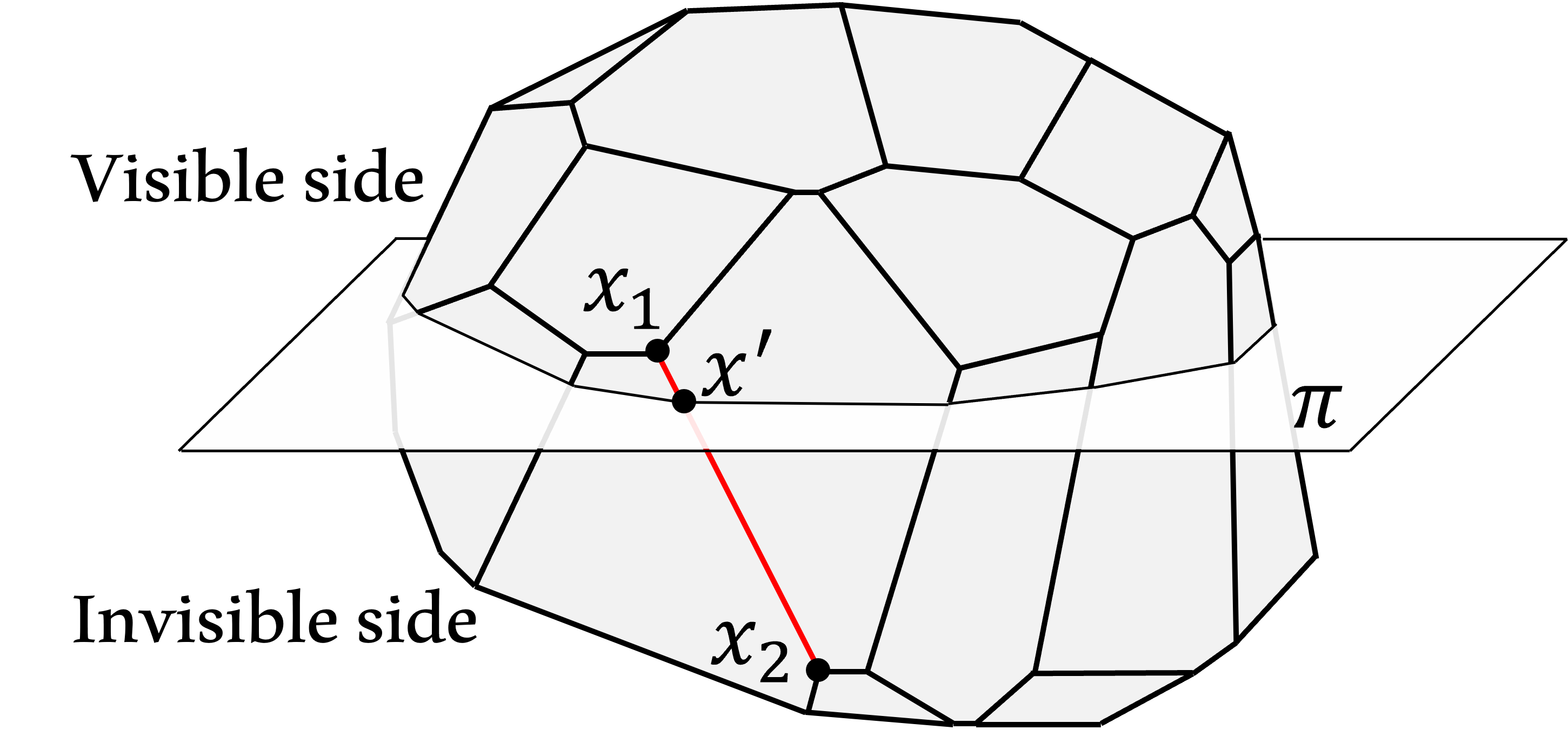}
\vspace{-8pt}
\end{wrapfigure}
For each corner of the convex polytope,
we keep the coordinates, as well as the three half-planes that define the corner.
For each edge
of the convex polytope,
we keep the identities of the two endpoints. 
As the inset figure shows, when a new half-plane~$\pi$ comes, we remove the vertices and the edges lying on the invisible side of~$\pi$.
If $\pi$ intersects a surviving edge $x_1x_2$ at a new point $x'$,
then $x'$ defines a new corner point 
and replaces the invisible endpoint of $x_1x_2$ at the same time. 

When all the half-planes are handled, 
the surviving vertices~$\{x_i\}_{i=1}^k$ are reported,
facilitating the inspection of interception;
See Theorem~\ref{thm:convexity:e}
and Theorem~\ref{thm:convexity:f}. 
Statistics show that the average time for 
computing $\textit{ConvexPoly}(v,e)$ (or $\textit{ConvexPoly}(v,f)$) is about 12 microseconds on the 20K-face Camel model, 
which is faster than the implementation in CGAL.

\subsection{Inspection in a flooding fashion}
\begin{algorithm}[htb]
\caption{Flooding inspection of $e$'s interceptors}
\label{alg:flooding}
\LinesNumbered
\KwIn{$e=v_1v_2$ and $\mathcal{V}_V$.}
Initialize an interceptor queue $Q=\{v_1,v_2\}$\;
\While{$Q$ is not empty}{
    Pop the front vertex $v_i$\;
   \If{$v_i$ intercepts $e$}{
    Update the interception table by taking~$v_i$ as the interceptor of~$e$\;
    \For {the neighboring generator $v_j$ (referring to $\mathcal{V}_V$)}{
    \If{$v_j$ has not been in $Q$}{
        Push $v_j$  into $Q$\;
    }
    }
    }
}
\vspace{-1mm}
\end{algorithm}

Suppose that the Voronoi diagram~$\mathcal{V}_V$
of the vertex set~$V$ has been precomputed. 
Rather than exhaustively 
inspect the interception between each vertex and each edge (or face),
we propose to perform inspection in a flooding fashion.
Initially, we build an empty interception table~$\mathcal{T}$. 
For a mesh edge~$e$, 
we inspect the vertices in~$V$ from $e$'s endpoints in a flooding fashion, where the neighboring relationship is defined by~$\mathcal{V}_V$.
Likewise, for each face~$f$,
we perform flooding inspection from the three vertices of~$f$.
All the detected interception pairs are kept in~$\mathcal{T}$. 
We summarize the flooding algorithm 
for inspection of $e$'s interceptors
in Algorithm~\ref{alg:flooding}.
We give a theorem on verifying the correctness of the flooding inspection scheme without missing any interception.

\begin{theorem} \label{thm:flood}
The flooding strategy of Algorithm~\ref{alg:flooding} ensures that all the interception pairs can be found.
\end{theorem}

\begin{figure}[htb]
    \centering
     \includegraphics[width=0.75\columnwidth]{"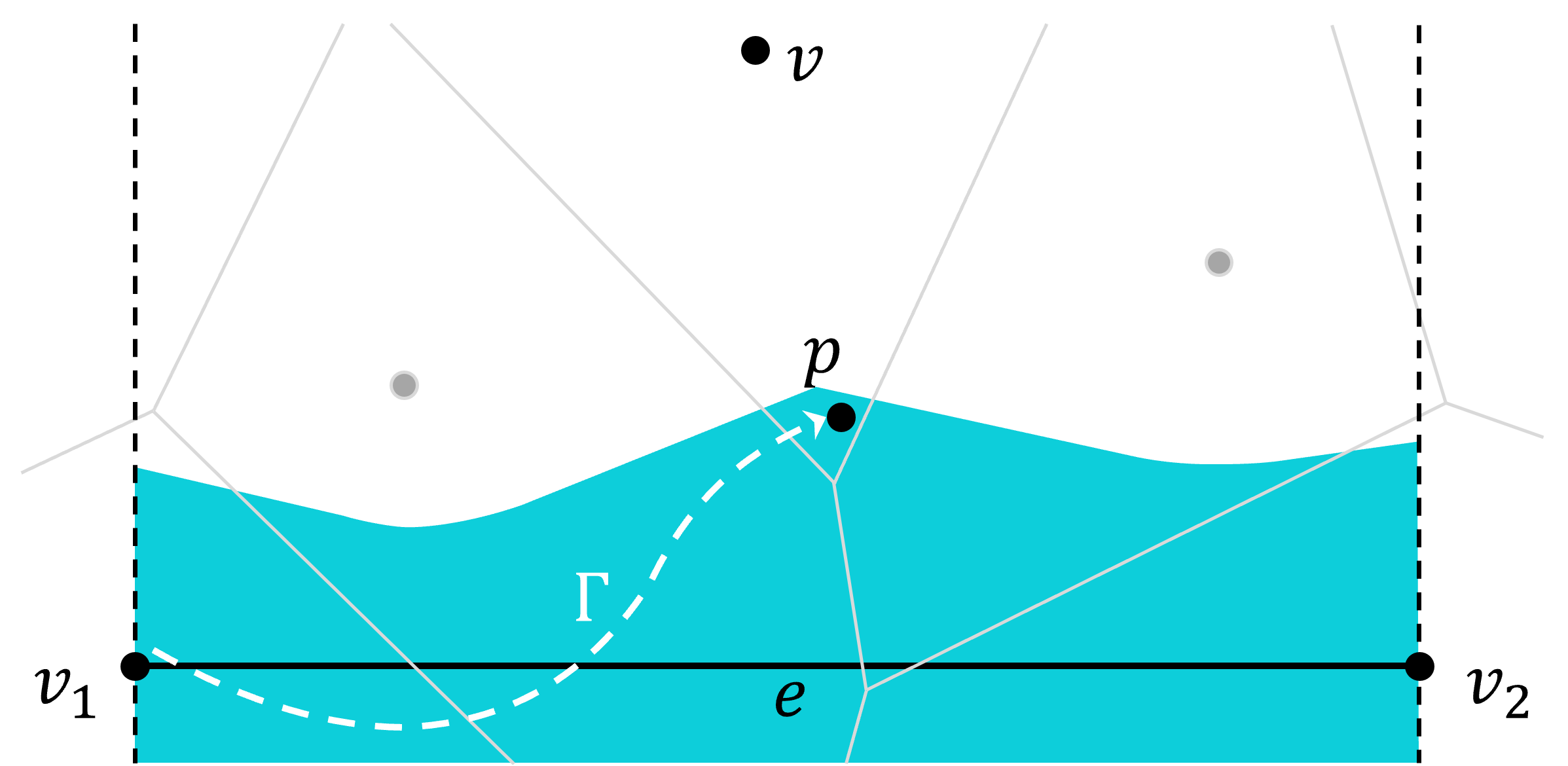"}
    \caption{Proof of Theorem~\ref{thm:flood}.
        $\textit{Cell}(e;\mathcal{V}_{V,E,F})$ of~$e=v_1v_2$ is colored in aqua blue.
        Each of $e$'s interceptors 
        can be found by the from-neighbor-to-neighbor flooding scheme.}
    \label{fig:flood}
    \vspace{-3mm}
\end{figure}

\begin{proof}
Suppose that $v$ is an interceptor of~$e=v_1v_2$.
We come to prove that $v$ must be found in the flooding process. 
Our proof is based on the fact that~$\textit{Cell}(e;\mathcal{V}_{V,E,F})$ 
is a connected region, which can be proved by contradiction (we ignore the proof).

As $v$ is the interceptor of $e=v_1v_2$, there must \new{be} at least one point, say, $p$, such that
$$
p \in \textit{Cell}(v;\mathcal{V}_V)\cap\textit{Cell}(e;\mathcal{V}_{V,E,F}).
$$
The connectedness of $\textit{Cell}(e;\mathcal{V}_{V,E,F})$
implies 
that there is 
a path $\Gamma\in\textit{Cell}(e;\mathcal{V}_{V,E,F})$
between $v_1$ and $p$ 
such that $\Gamma\backslash v_1$ is totally inside $\textit{Cell}(e;\mathcal{V}_{V,E,F})$.
Suppose that $\Gamma$ crosses 
a sequence of Voronoi \new{cells} in $\mathcal{V}_{V}$, denoted by \new{$\mathcal{C}$}. 
It is easy to verify the following two facts.
First, $v_1$ is a natural interceptor of $e$.
Second, every Voronoi cell that has intersections with $\Gamma$ gives an interceptor for $e$.
Therefore,
we can take $v_1$ as the first interceptor,
and then trace the other interceptors 
along \new{$\mathcal{C}$} 
until $v$ is found. 
\end{proof}

\subsection{Further optimization in the query phase} \label{subsec:Further:optimization}
In the query stage,
we first find the nearest vertex~$v$ by KDT search, 
followed by looking up the interception table.
A na\"{i}ve strategy is to enumerate 
all the primitives in the interception table
to identify the one that is closest to the query point~$q$. 
However, the query becomes inefficient when the interception table is very long.
Recall that
an edge $e$ is said to be intercepted by $v$ 
if $\textit{Cell}(v;\mathcal{V}_{V})\cap\textit{Cell}(e;\mathcal{V}_{V,E,F})\neq \emptyset$, 
regardless of the location of the specific query point~$q$.

\begin{figure}[htb]
    \centering
     \includegraphics[width=0.6\columnwidth]{"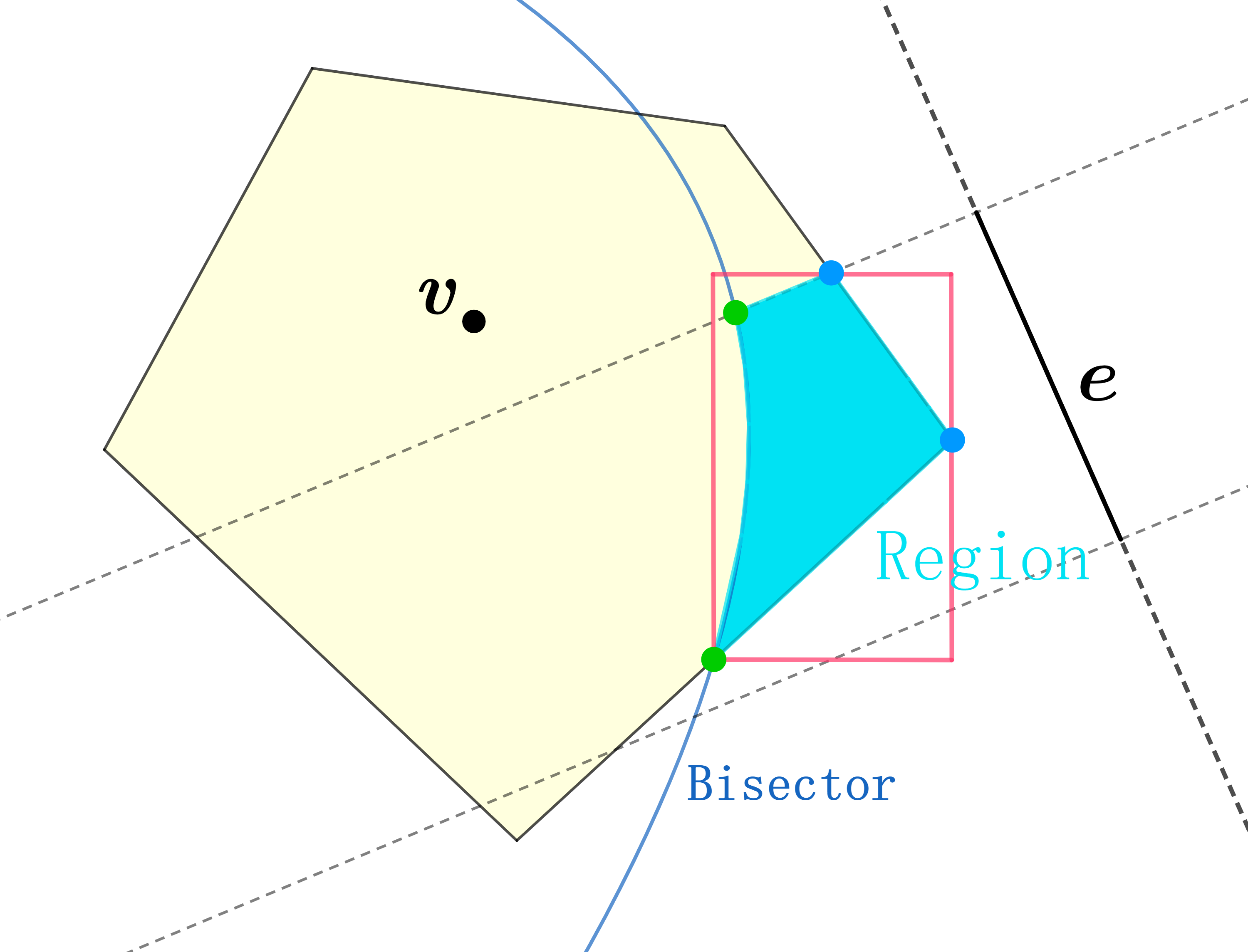"}
    \caption{
    An illustration of $\textit{Region}(v,e)$ (colored in aqua blue) as well as its bounding box (colored in red).}
    \label{fig:MBR}
    \vspace{-3mm}
\end{figure}
Based on the discussion in Section~\ref{subsec:interceptionfilter}, 
the assertion $\textit{Cell}(v;\mathcal{V}_{V})\cap\textit{Cell}(e;\mathcal{V}_{V,E,F})\neq \emptyset$
can be relaxed to $$\textit{ConvexPoly}(v,e)\cap \textit{Bisect}^e(v,l_e)\neq \emptyset;$$
See Figure~\ref{fig:MBR}.
We denote $\textit{ConvexPoly}(v,e)\cap \textit{Bisect}^e(v,l_e)$ by $\textit{Region}(v,e)$.
To this end, whether the edge~$e$
 contributes to the minimum distance 
can be reduced to check if~$q$ is located in~$\textit{Region}(v,e)$.
However, $\textit{Region}(v,e)$ is generally non-convex,
making it \new{difficult} to perform the inside-outside test.

Therefore, for each edge \new{$e$ (resp., face $f$)} in the interception list of~$v$, 
we suggest enclosing $\textit{Region}(v,e)$ \new{(resp., $\textit{Region}(v,f)$)} by its bounding box and then organizing the bounding boxes into an R-tree, 
\new{finally producing one R-tree per interception list. }

Besides, 
when determining whether a primitive, say, $e$ (resp., $f$),
can provide the minimum distance,
we first check whether $q\in\textit{Space}^\perp(e)$ (resp., $q\in\textit{Space}^\perp(f)$) or not.
Only when the assertion is true,
we come to calculate the distance from $q$ to the straight line of $e$ 
(resp., the plane of $f$).

To summarize, 
after the nearest vertex~$v$ is found by KDT search,  
the geometric primitives are further filtered out by R-tree.
For the surviving geometric primitives,
we conduct an exhaustive comparison to accomplish the point-to-mesh distance query.

\section{Evaluation}
We conducted experiments on a PC with AMD Ryzen 9 5950X 16-core processor.
Our implementation is written in C++.
We call TetGen (version 1.6.0)~\cite{Si2015TetGenAD} to compute the Voronoi diagram w.r.t.~the vertex set~$V$.
We first give the performance statistics of the proposed algorithm.
After that, we compare our algorithm with PQP\footnote{\new{The modified PQP version includes a direct interface for point-to-mesh distance queries, and the code is available at \url{https://mewangcl.github.io/Projects/MeshThickeningProj.htm}.}} and FCPW in terms of various indicators.

Note that FCPW supports SIMD parallelism and we set the CPU based SIMD width to 4 in our experiments. 
There is no parallelism in PQP and our code. 
For each test model, we randomly sample a million query points within the 10x axis-aligned bounding box.

\begin{figure}[ht]
    \centering
    \includegraphics[width=\columnwidth]{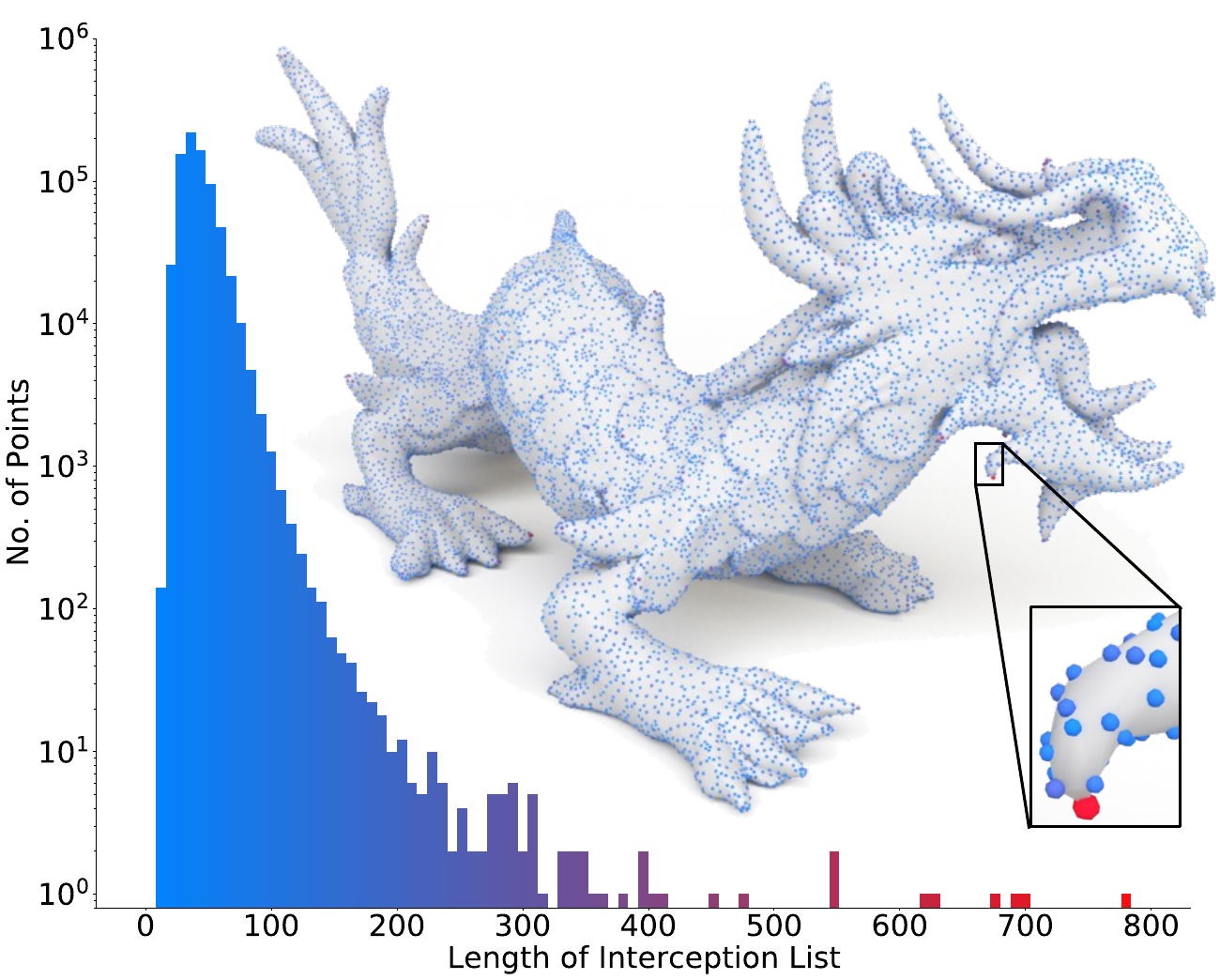}
    \caption{Different vertices have different numbers of 
intercepted primitives. 
We visualize the vertices in varying colors according to the length of interception list. 
We set the vertical axis to be in a logarithmic scale. 
The vertex colored in red owns a long interception list.
   }
    \label{fig:interception_dragon}
\end{figure}

\subsection{Interception list}

\paragraph{Length of the interception list for a vertex}
Recall that in the interception table, we keep the intercepted geometric primitives, i.e., 
edges and faces, for each vertex in~$V$.
As different vertices have different numbers of 
intercepted primitives, 
we take the 1500K-face Dragon model as the input, shown in Figure~\ref{fig:interception_dragon}, to observe how the length of the interception list of a vertex varies with the position on the surface (the horizontal axis: the length of the interception list; the vertical axis: the number of vertices). 
It can be seen that for most of the vertices,
the length of the interception list of a vertex ranges from 20 to 100. The average length for this example is about 41.
However, 
there is an occurrence that the interception list is very long.
In Figure~\ref{fig:interception_dragon},
the vertex colored with a red dot keeps 782 intercepted primitives.
Once the vertex is retrieved by the KDT, 
it is time-consuming to exhaust every intercepted primitive 
in its interception list.
That's why we introduce an R-tree based filtering method (see Section~\ref{subsec:Further:optimization}) to reduce
the number of candidate primitives. 

\begin{figure}[ht]
    \centering
    \includegraphics[width=\linewidth]{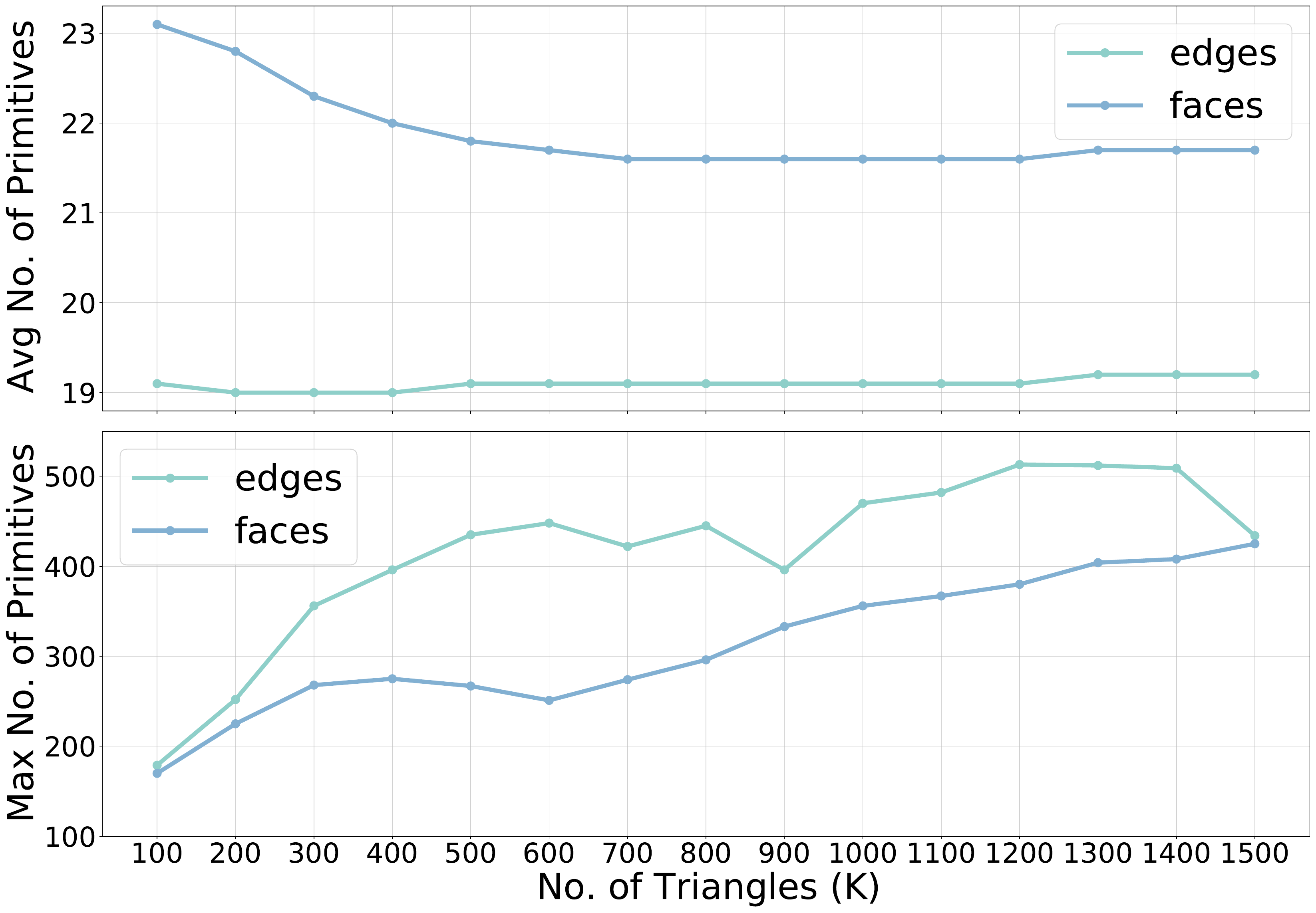}
    \caption{The maximum and average number of intercepted primitives of a vertex w.r.t. the mesh resolution of the  Dragon model.}
    \label{fig:invade_cnt}
    \vspace{-3mm}
\end{figure}

We simplify the Dragon model into 100K, 200K, $\cdots$, 1500K faces respectively and then record the average and maximum number of the intercepted primitives for a vertex. 
Statistics in Figure~\ref{fig:invade_cnt} show that the average length of interception list remains nearly unchanged with the increase of mesh resolution, but there is a slight rise in the maximum length.

\begin{table}[htb]
\caption{The average (Avg$^1$), size-weighted average (Avg$^2$) and maximum numbers of intercepted primitives on 5 models,
where ``size-weighted'' means that
an interception list is weighed by the volume of the Voronoi cell of its interceptor vertex. The average and maximum numbers of tests for one query operation is also reported.  \new{Note that we use two sets of numbers for each cell to indicate respectively the number of edges and the number of faces.} }
\centering
\resizebox{\columnwidth}{!}{%
\begin{tabular}{ccccccccccc}
\hline
            & &             &   & \multicolumn{3}{c}{Intercepted (edges, faces)}  &  & \multicolumn{2}{c}{Tested (edges, faces)}\\ \cmidrule{5-7}\cmidrule{9-10}
Model       & &Faces        &   & Avg$^1$        & Avg$^2$      & Max                       &  & Avg                     & Max            \\  \cline{1-1}\cline{3-3}\cline{5-7}\cline{9-10}
Camel       & & 19510       &   & 18.0, 26.3         & 15.0, 17.2         & 105, 130                  &  & 5.5, 2.7              & 12, 12           \\
Armadillo   & & 99976       &   & 19.0, 23.6         & 17.1, 20.7         & 81, 91                    &  & 4.7, 1.9              & 13, 13           \\
Sponza      & & 262196      &   & 21.4, 26.8         & 34.8, 29.4         & 296, 168                  &  & 0.4, 0.1              & 23, 28           \\
Lucy        & & 525814      &   & 18.3, 23.7         & 23.1, 22.6         & 346, 232                  &  & 4.3, 1.6              & 14, 9            \\
Dragon      & & 1499852     &   & 19.2, 21.7         & 16.6, 19.2         & 434, 425                  &  & 5.1, 2.0              & 13, 13           \\ \hline
\end{tabular}%
}
\label{tab:data2}
\end{table}

\paragraph{Number of tested
primitives}
We must point out that different vertices have different chances of being accessed, depending on
the size of the Voronoi cell~$\textit{Cell}(v;\mathcal{V}_V)$. 
Therefore, we take the size of $\textit{Cell}(v;\mathcal{V}_V)$
to weigh $v$'s access chance.

As Table~\ref{tab:data2} shows, the size-weighted average number (Avg$^2$) is slightly different from the non-weighted average number (Avg$^1$).

Recall that we propose an R-tree based filtering technique 
to filter out most of the primitives that do not help. 
It can be seen from the right side of Table~\ref{tab:data2}
that both the average and maximum numbers of tested primitives,
for each query, are significantly reduced. 
To summarize, the filtering technique is helpful in improving the overall performance, especially when there exist long interception lists. \new{A more detailed discussion about the speed-up gain in the query performance by R-tree can be found in Section~\ref{sec:query_perform}.}

\subsection{Preprocessing cost}
\paragraph{Cost breakdown}
The overall preprocessing cost
consists of four parts, i.e.,
1)~KDT construction,
2)~Voronoi diagram generation,\\
3)~flood-based interception inspection 
and 
4)~\new{R-tree construction}.

The cost breakdown generalizes to the 5 models (\new{see Table~\ref{tab:piechart}}) as well as other tested models in large datasets.

It can be seen that 
the computation of the interception table 
takes about 90\% of the total preprocessing time.

Despite this, 
the flooding scheme is still very helpful.
If we inspect interception without flooding scheme, which requires to test every vertex-edge pair and every vertex-face pair,
this brute-force manner requires more than 24 hours 
to accomplish the task on the 1500K-face Dragon model. 
The cost of interception inspection is 
reduced to 2 minutes with the help of 
the flooding scheme.

\begin{table}[htb]
\caption{
    \new{The computational time consumption and corresponding proportion about KDT construction, Voronoi diagram generation, interception inspection and R-tree construction in preprocessing procedure.}
    }
\centering
\resizebox{\columnwidth}{!}{%
\begin{tabular}{ccccccccccccc}
\hline
         & & \multicolumn{2}{c}{\makecell{KDT\\construction}}& & \multicolumn{2}{c}{\makecell{Voronoi\\diagram}}& & \multicolumn{2}{c}{\makecell{Interception\\inspection}}& & \multicolumn{2}{c}{\makecell{R-tree\\construction}} \\ \cmidrule{3-4}\cmidrule{6-7}\cmidrule{9-10}\cmidrule{12-13}
Model    & & T($ms$)         & Prop.        & & T($ms$)          & Prop.        & & T($ms$)              & Prop.            & & T($ms$)           & Prop.          \\ \cline{1-1}\cline{3-4}\cline{6-7}\cline{9-10}\cline{12-13}
Camel    & & 1.9             & 0.2\%              & & 105.4            & 10.6\%             & & 854.2                  & 85.7\%                   & & 17.1              & 1.7\%                \\
Armadillo& & 9.4             & 0.2\%              & & 718.8            & 15.0\%             & & 3847.5                 & 80.4\%                   & & 65.3              & 1.4\%                \\
Sponza   & & 24.9            & 0.2\%              & & 1936.2           & 12.3\%             & & 13245.3                & 84.3\%                   & & 232.1             & 1.5\%                \\
Lucy     & & 53.1            & 0.2\%              & & 4100.0           & 16.6\%             & & 19436.6                & 78.9\%                   & & 413.2             & 1.7\%                \\
Dragon   & & 154.5           & 0.2\%              & & 12163.0          & 12.6\%             & & 80596.8                & 83.7\%                   & & 1093.9            & 1.1\%               \\
\hline
\end{tabular}
}
    \label{tab:piechart}
\end{table}

\begin{figure}[ht]
    \centering
    \includegraphics[width=\linewidth]{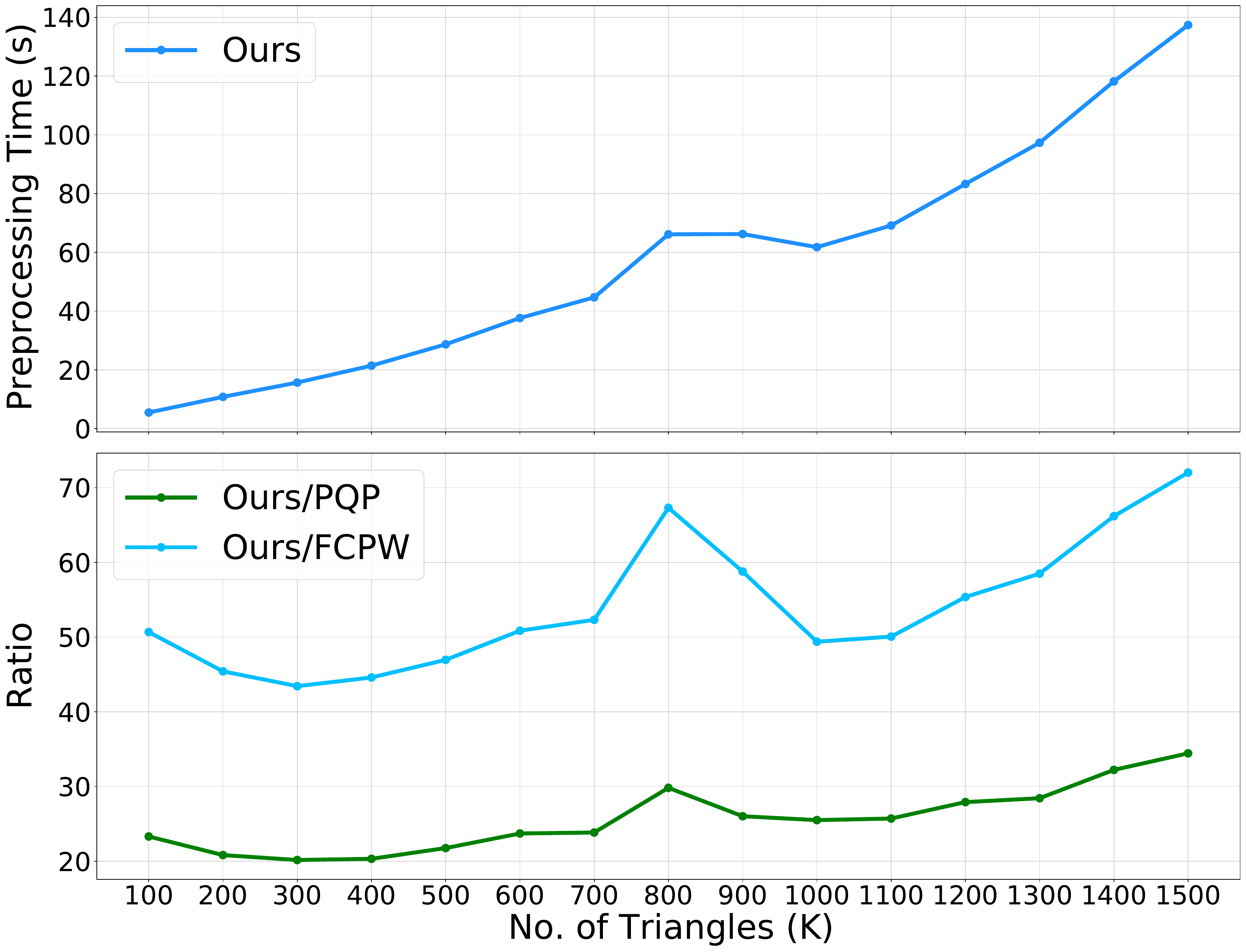}
    \caption{The preprocessing cost on the Dragon model with varying resolutions. 
    Top: our preprocessing cost v.s. the mesh resolution. Bottom: the comparison about the preprocessing cost
    among PQP, FCPW and ours. 
    }
    \label{fig:dragon:build}
    \vspace{-3mm}
\end{figure}

\begin{figure}[ht]
    \centering
    \includegraphics[width=\linewidth]{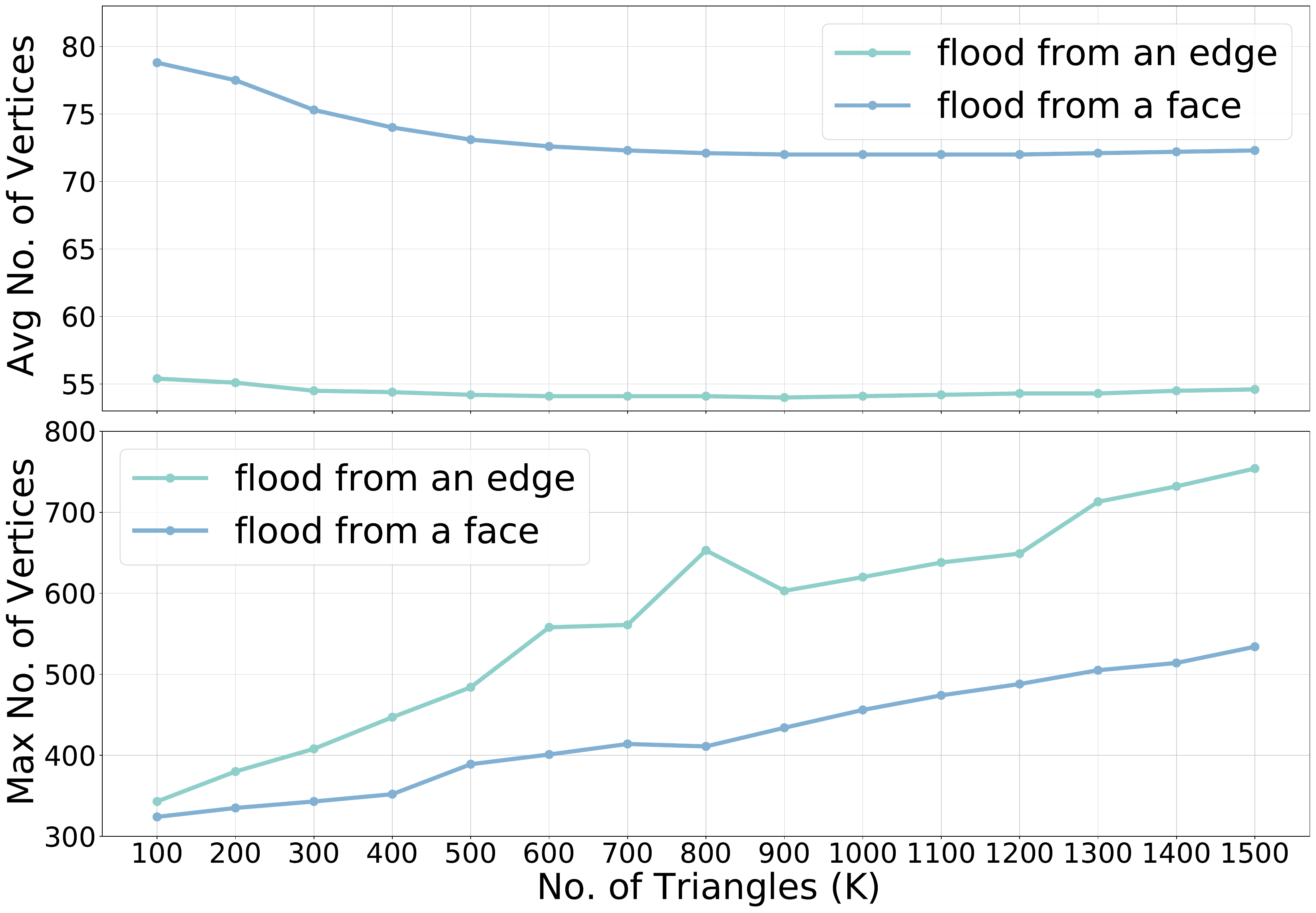}
    \caption{The average and maximum numbers of examined vertices when flooding from an edge or a face.
    The tests are made on the Dragon model with different  resolutions.}
    \label{fig:flood_cnt}
    \vspace{-3mm}
\end{figure}

In Figure~\ref{fig:dragon:build}, we give the statistics about the preprocessing time on the Dragon model with varying resolutions.
Although the worst-case theoretical time complexity of flooding is $O(nm)$ ($n$ is the number of vertices and $m$ 
is the total number of edges and faces),
the empirical time complexity is quasilinear w.r.t. $n$. 
Furthermore, 
it can be seen from Figure~\ref{fig:flood_cnt}
that the average number
of examined vertices during flooding remains almost unchanged
while the maximum number has a slight rise.

\paragraph{Comparison with existing libraries}
Our algorithm requires a much larger preprocessing cost
than PQP and FCPW.
For example, PQP and FCPW take about 4.0 seconds and 1.9 seconds respectively to construct the BVH,
but our algorithm 
requires 137.4 seconds for preprocessing the 1500K-face Dragon model.
By using the Dragon model as the test,
we provide the comparison about the preprocessing cost among PQP, FCPW
and ours in Figure~\ref{fig:dragon:build}.
We further compare them on the Thingi10K dataset;
See the statistics of the preprocessing costs
in Figure~\ref{fig:Thingi10K_build}.

\subsection{Query performance}\label{sec:query_perform}
\begin{figure}[ht]
    \centering
    \includegraphics[width=\columnwidth]{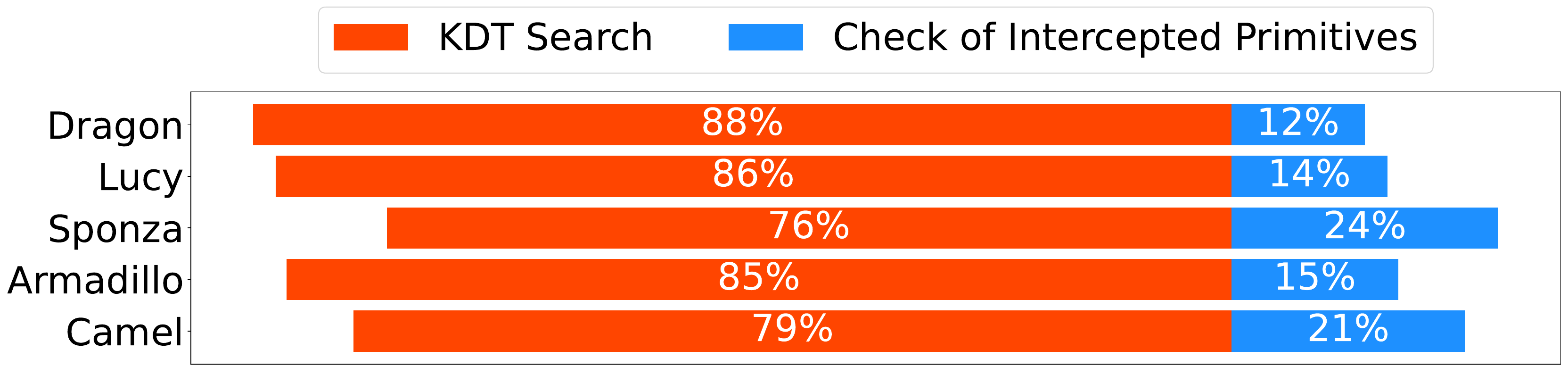}
    \caption{Cost breakdown of query performance on 5 models.}
    \label{fig:query_breakdown}
    \vspace{-3mm}
\end{figure}

\paragraph{Cost breakdown}
The query stage of our algorithm involves
two operations: (1)~finding the nearest vertex~$v$ by KDT search,
and (2)~identifying the closest geometric primitive by visiting $v$'s interception list. 
The average timing costs of the two parts on the 1500K-face Dragon model are respectively 2.83 microseconds and 0.39 microseconds.
We further plot the cost breakdown in Figure~\ref{fig:query_breakdown}. 
It can be seen that
KDT search takes over half of the total query time,
which shows the effectiveness of our R-tree based filtering rule.

\begin{figure}[ht]
    \centering
    \includegraphics[width=\linewidth]{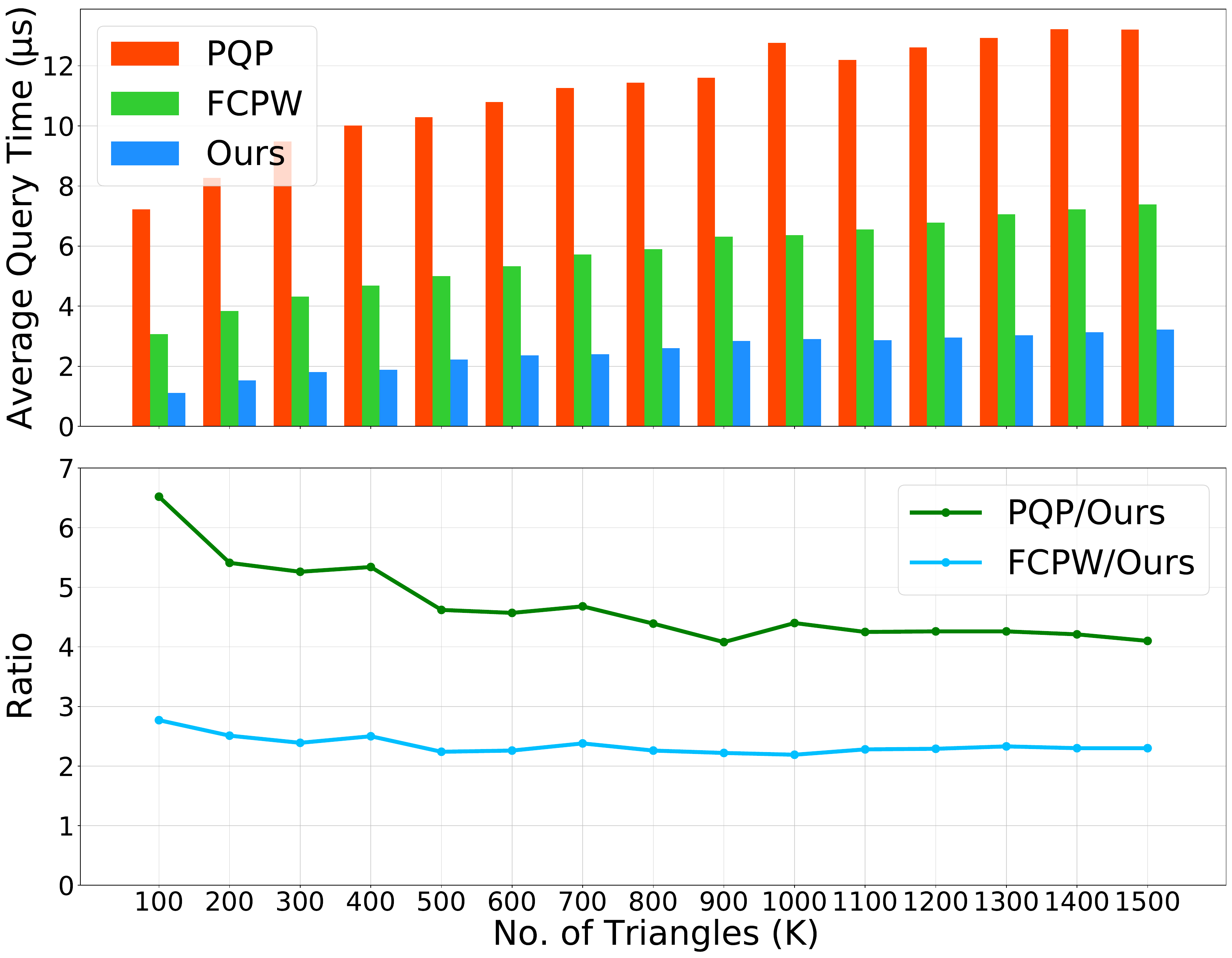}
    \caption{Comparison about query performance on the Dragon model with varying resolutions. Top: the average timing cost per query for PQP, FCPW and ours ($\mu$s). Bottom:  
    the comparison about the query cost
    among PQP, FCPW and ours. 
    }
    \label{fig:dragon:query}
    \vspace{-3mm}
\end{figure}

\begin{figure}[ht]
    \centering
    \includegraphics[width=\columnwidth]{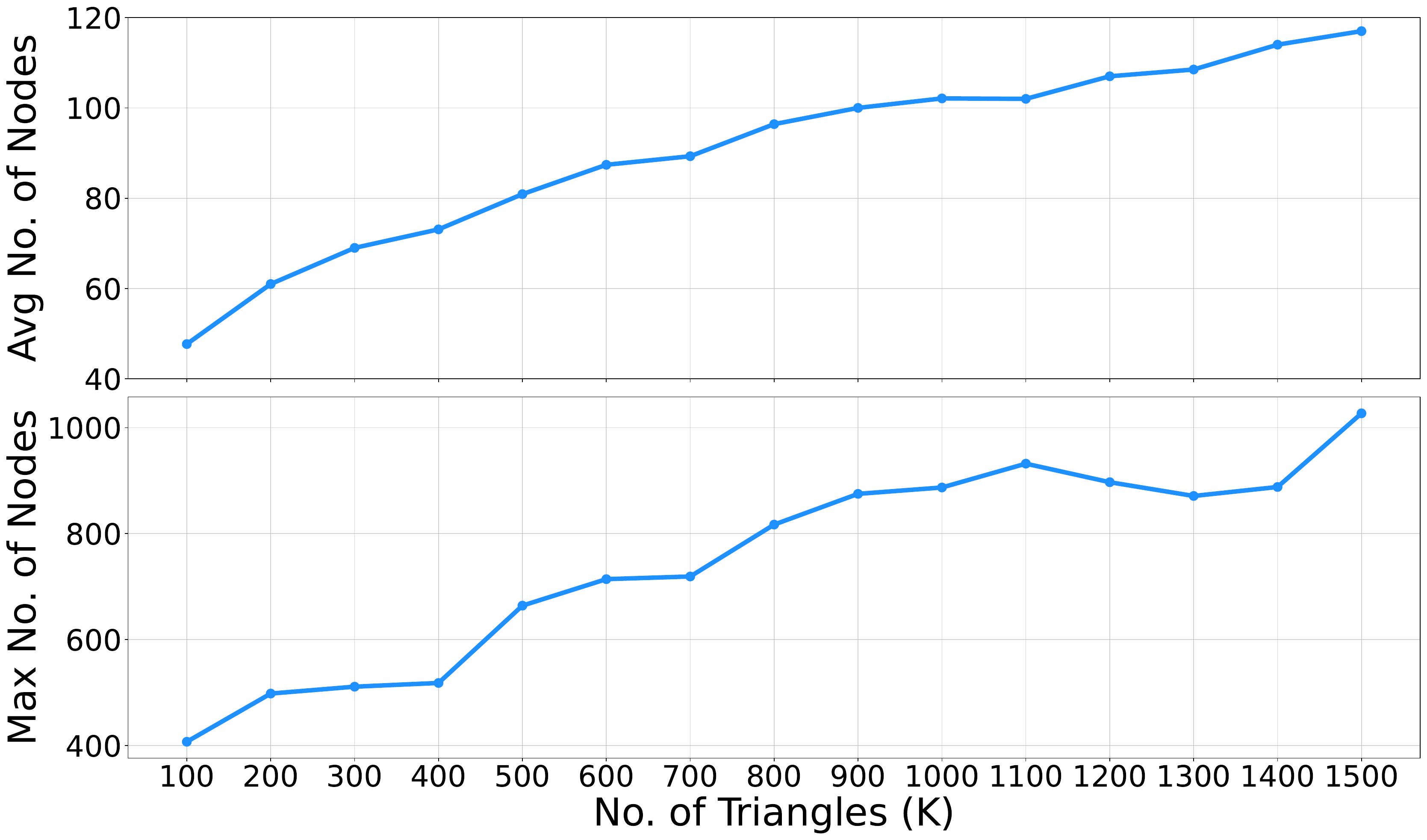}
    \caption{The average and maximum numbers of examined tree nodes during KDT search w.r.t. mesh resolution of the Dragon model.}
    \label{fig:kdt_nodes}
    \vspace{-3mm}
\end{figure}

Likewise, we plot the average query cost w.r.t. mesh resolution of the Dragon model in Figure~\ref{fig:dragon:query}. 
It can be seen that our average query cost climbs gently with the increasing number of faces. 
In Figure~\ref{fig:kdt_nodes}, we further provide the average and maximum numbers of examined tree nodes during KDT search.

\paragraph{Comparison with existing libraries}
Figure~\ref{fig:dragon:query} gives a plot for visualizing 
 the comparison about the query cost
    among PQP, FCPW and ours. 

It can be seen that
our algorithm runs at least 4 times as fast as PQP and 2 times faster than FCPW even on the model with 1500K faces, which shows that our algorithm has a better query performance,
especially on large-sized 3D models.
Figure~\ref{fig:Thingi10K_query} gives more comprehensive 
statistics about the query performance.

\begin{figure}[ht]
    \centering
    \includegraphics[width=\linewidth]{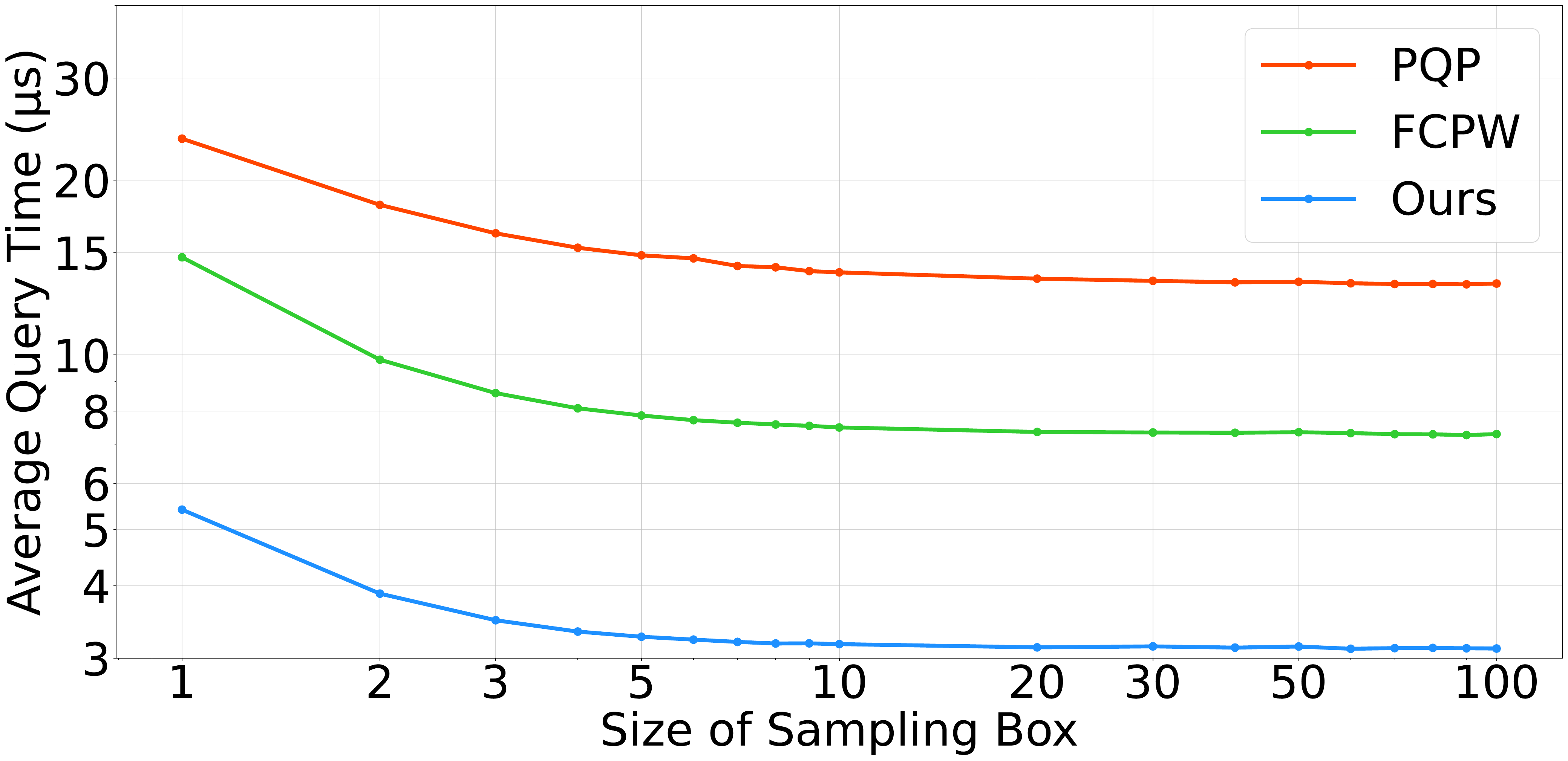}
    \caption{Query performance w.r.t. the size of sampling box of query points on the Dragon model. (log-log plot)}
    \label{fig:query_box}
    \vspace{-3mm}
\end{figure}

\paragraph{
Near and far query points}
It is necessary to make clear
how the query performance depends on the distance between the query point and the surface.
In Figure~\ref{fig:query_box},
we plot the dependence of PQP's, FCPW's and our query performance 
on the distance,
where the horizontal axis indicates  
\new{the ratio of the maximum side length of sampling box to that of the minimum bounding box}. 
It can be seen that, like PQP and FCPW,  
our query becomes more time-efficient 
with the increasing distance between the query point and the surface.

\begin{figure}[ht]
    \centering
    \includegraphics[width=\linewidth]{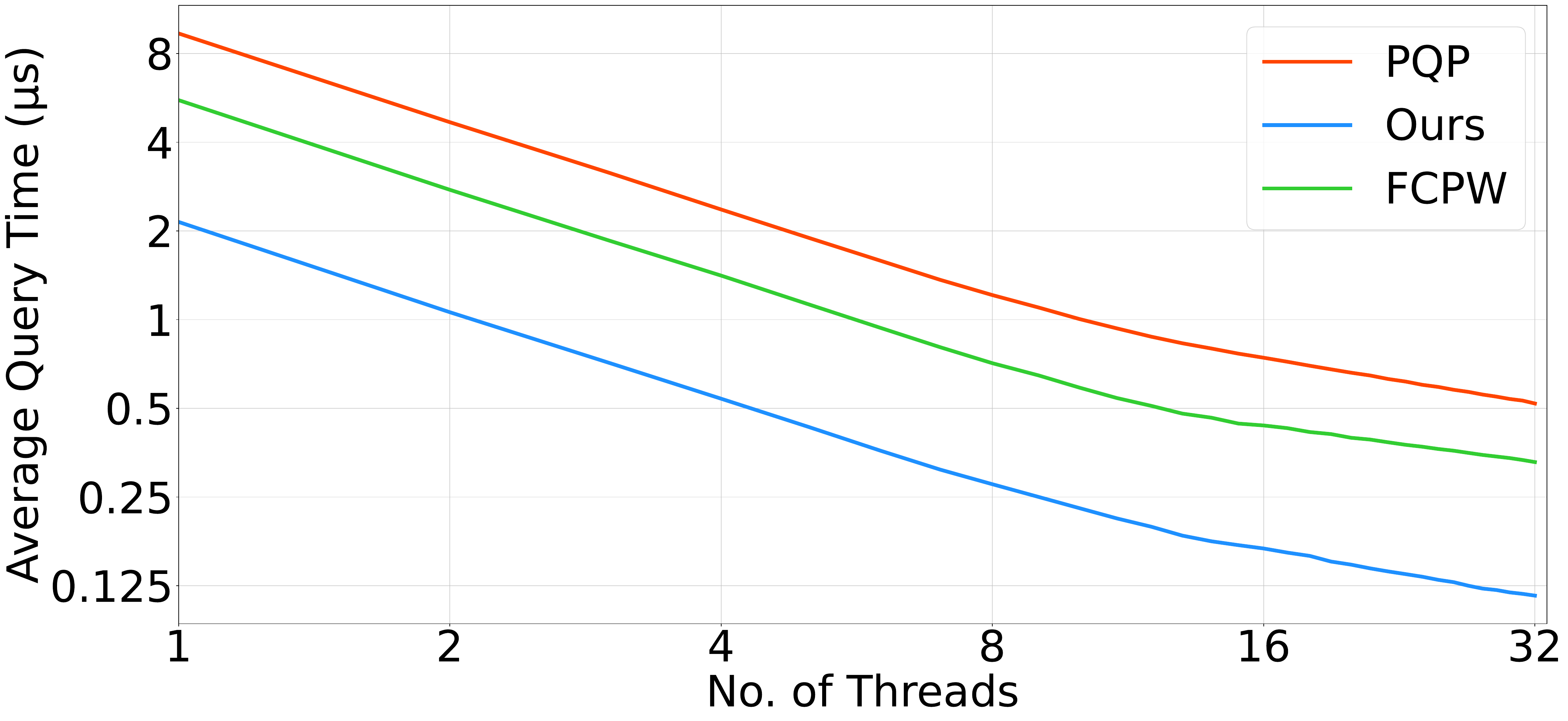}
    \caption{Comparison about average timing cost among PQP, FCPW and our algorithm under multi-thread parallelism. (log-log plot)
    }
    \label{fig:parallel_tbb}
    \vspace{-3mm}
\end{figure}

\paragraph{Thread-based parallelism}
We perform thread-based parallelism tests, which simply divides independent queries into groups and distribute them 
to separate threads.
As Figure~\ref{fig:parallel_tbb}
shows,
our algorithm runs much faster than PQP and FCPW 
under the same thread-based parallelism and the parallel performance is close to linear.

\new{
\paragraph{Speed-up gain of R-tree}
The use of R-tree 
does help when the interception list is very long.
Compared with the brute-force strategy of testing all primitives in the interception list to find the closest point, the R-tree-based method can effectively filter out unnecessary primitives. 
Table~\ref{tab:r-tree_acce} shows some examples. For general models without long interception lists, such as Camel and Armadillo, the filtering effect of R-tree is not significant.
}

\begin{table}[htb]
\caption{
    \new{Comparison about the query cost among PQP ($T_{PQP}$), FCPW ($T_{FCPW}$) and our approaches with ($T_{R}$) or without the use of R-tree ($T_{BF}$).
    The experiments are conducted on totally 10 models, where the last 5 models, selected from the Thingi10K dataset,
    have long interception lists. Avg$^1$ and Avg$^2$ indicate the average and size-weighted average numbers of intercepted primitives respectively.
    The speed-up gain of R-tree is conspicuous for models with long interception lists.}
    }
\centering
\resizebox{\columnwidth}{!}{%
\begin{tabular}{cccccccccccccc}
\hline
         & & \multicolumn{2}{c}{\makecell{Tested (edges, faces)}}\\ \cmidrule{3-4}
Model    & & Avg$^1$              & Avg$^2$     & & $T_{PQP}$($\mu s$)    & & $T_{FCPW}$($\mu s$)       & & $T_{BF}$($\mu s$)   & & $T_{R}$($\mu s$)        & & $\frac{T_{BF}}{T_{R}}$ \\ 
\cline{1-1}\cline{3-4}\cline{6-6}\cline{8-8}\cline{10-10}\cline{12-12}\cline{14-14}
Camel    & & 15.0, 17.2           & 5.5, 2.7     & & 7.47    & & 3.64      & & 1.43            & & 1.28             & & 1.12  \\
Armadillo& & 17.1, 20.7           & 4.7, 1.9     & & 9.64    & & 5.43      & & 2.45            & & 2.19             & & 1.11  \\
Sponza   & & 34.8, 29.4           & 0.4, 0.1     & & 7.69    & & 0.79      & & 0.57            & & 0.41             & & 1.36  \\
Lucy     & & 23.1, 22.6           & 4.3, 1.6     & & 11.84   & & 5.43      & & 2.44            & & 2.13             & & 1.14 \\
Dragon   & & 16.6, 19.2           & 5.1, 2.0     & & 13.83   & & 7.56      & & 3.54            & & 3.34             & & 1.06 \\
\#378036 & & 383.5, 385.7         & 6.8, 2.6     & & 37.27   & & 20.04     & & 15.77           & & 7.60             & & 2.07 \\
\#69078  & & 356.5, 4981.7        & 5.2, 6.2     & & 9.45    & & 6.79      & & 60.54           & & 3.43             & & 17.66 \\
\#82324   & & 743.6, 817.3        & 1.5, 0.8     & & 5.18    & & 1.12      & & 13.43           & & 0.54             & & 25.06 \\
\#1472696 & & 2646.6 1971.6       & 0.4, 0.1     & & 6.94    & & 0.51      & & 30.51           & & 0.47             & & 64.97 \\
\#236143  & & 2125.6, 1533.5      & 0.2, 0.1     & & 5.61    & & 0.79      & & 37.51           & & 0.37             & & 100.75 \\
\hline
\end{tabular}
}
    \label{tab:r-tree_acce}
\end{table}

\subsection{Memory usage}
The memory requirements consist of four parts, i.e.,
1)~KDT structure,
2)~geometric information of primitives,
3)~interception lists
and 4)~R-tree structures. 
For the Dragon model,
the four parts of memory usage are respectively 2\%, 10\%, 67\% and 21\%.
The proportions are similar for a general input model.
In contrast,
our memory consumption is generally larger than PQP and FCPW.
Taking the Dragon model for an example, \new{
PQP and FCPW require 0.61 GB and 0.16 GB of memory, respectively, while ours requires 1.78 GB. 
It is necessary to mention that FCPW uses single-precision 
variables
while PQP and ours use double-precision variables.
}
We give comprehensive statistics of memory consumption in Figure~\ref{fig:Thingi10K_memory} on Thingi10K.

\begin{figure}[ht]
    \centering
    \includegraphics[width=\columnwidth]{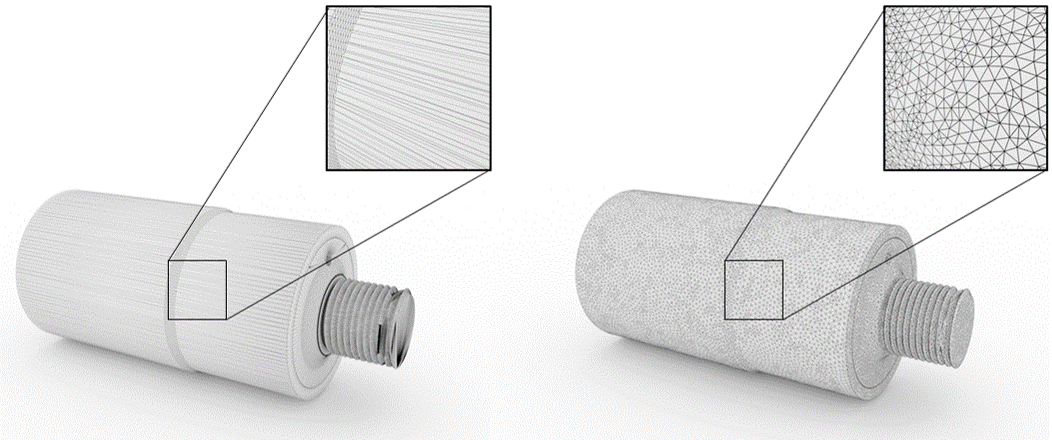}
    \makebox[0.49\columnwidth][c]{(a) Bad triangulation}
    \makebox[0.49\columnwidth][c]{(b) Good triangulation}
    \caption{
    The Water-Bottle model has 49,974 triangles. 
    (a) Low-quality mesh with many long skinny triangles. 
    (b) High-quality mesh with the same number of triangle facets.
   }
    \label{fig:mesh_quality}
\end{figure}

\begin{table}[ht]
\caption{Each of the six models
has a low-quality triangle mesh as well as a high-quality counterpart. 
We list the query performance of PQP, FCPW and ours on these test models. }
\centering
\resizebox{\columnwidth}{!}{
\begin{NiceTabular}{|c|c|c|p{0.8cm}<{\centering}|c|p{0.8cm}<{\centering}|c|p{0.9cm}<{\centering}|c|p{0.8cm}<{\centering}|c|}
\toprule
\multirow{2}{*}[-4pt]{Model}  & \multicolumn{2}{c}{Tri Quality} & \multicolumn{2}{c}{Bad-tri percent} & \multicolumn{2}{c}{PQP query ($\mu$s)} & \multicolumn{2}{c}{FCPW query ($\mu$s)} & \multicolumn{2}{c}{Our query ($\mu$s)}\\\cline{2-11}
& low & high & low & high & low & high & low & high & low & high\\
\midrule
\begin{minipage}[ht]{0.20\columnwidth}
  \centering
  \raisebox{-1.2\height}{\includegraphics[width=\linewidth]{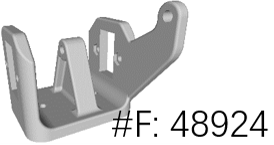}}
 \end{minipage}
 &0.387 &0.755 &32.96\% &0.13\% & 13.12 &9.41 &6.14 &3.68 &1.61 &1.31\\\hline
\begin{minipage}[ht]{0.20\columnwidth}
  \centering
  \raisebox{-1.2\height}{\includegraphics[width=\linewidth]{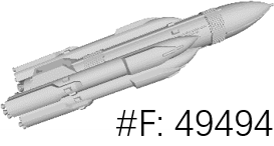}}
 \end{minipage} &0.344 &0.603 &44.47\% &7.02\% &15.20 & 6.98 &4.44 &3.52 &1.42 &1.18\\\hline
\begin{minipage}[ht]{0.20\columnwidth}
  \centering
  \raisebox{-1.2\height}{\includegraphics[width=\linewidth]{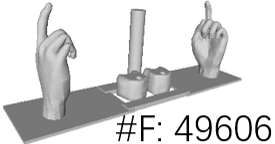}}
 \end{minipage} &0.598 &0.763 & 7.52\% &0.50\% & 6.50 & 6.52 &2.16 &2.02 &0.66 &0.65\\\hline
\begin{minipage}[ht]{0.20\columnwidth}
  \centering
  \raisebox{-1.2\height}{\includegraphics[width=\linewidth]{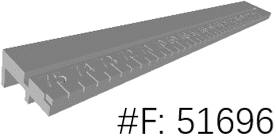}}
 \end{minipage} &0.061 &0.704 &87.75\% &1.28\% & 7.53 & 7.21 &0.76 &0.83 &0.31 &0.31\\\hline
\begin{minipage}[ht]{0.20\columnwidth}
  \centering
  \raisebox{-1.2\height}{\includegraphics[width=\linewidth]{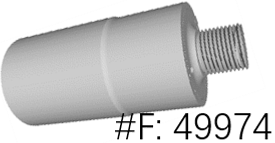}}
 \end{minipage} &0.286 &0.756 &59.19\% &3.29\% &15.98 & 8.66 &3.68 &5.22 &1.05 &1.95\\\hline
\begin{minipage}[ht]{0.20\columnwidth}
  \centering
  \raisebox{-1.2\height}{\includegraphics[width=\linewidth]{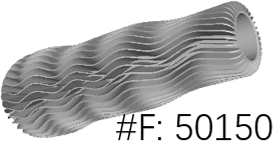}}
 \end{minipage} &0.292 &0.621 &58.39\% &3.17\% &24.96 &15.02 &11.24 &5.54 &2.72 &1.66\\
\bottomrule
\end{NiceTabular}
}
\label{tab:mehs_quality}
\vspace{-3mm}
\end{table}

\subsection{Extreme tests}
\paragraph{Poor triangulation quality}
\new{Generally speaking, irregularly shaped triangle faces would cause difficulties in the node splitting procedure of BVH structure generation and affect the performance of closest point query.}
It is necessary to observe if poor triangulation slows down the query performance of our algorithm.
We select six poorly-triangulated models from the Thingi10K dataset
and for each of them, we use the mesh optimization tool~\cite{hoppe1993mesh}
to get a high-quality counterpart with the same number of triangles. 
The quality of a triangle~$t$ can be measured by
$$
Q(t) = \frac{6}{\sqrt{3}}\frac{S_t}{{p_t}{h_t}},
$$
where $S_t, p_t, h_t$ are respectively the area, the half-perimeter of triangle and the longest edge length. $Q(t)$ ranges from 0 to 1, and equals~1 when~$t$ is a regular triangle.
Table~\ref{tab:mehs_quality} gives 
the overall triangle quality as well as the ratio of bad triangles. 
Here a triangle is considered to ``bad'' if the minimum angle is less than 10 degrees. 
Statistics show that our query performance does not have a conspicuous drop
when the triangulation quality is diminished. 
Taking the Water-Bottle model (as shown in Figure~\ref{fig:mesh_quality})
for an example, our average query cost is 1.95$\mu$s on the high-quality model
while the query cost becomes 1.05$\mu$s on the poorly-triangulated model.
In contrast, PQP is sensitive to triangle quality, 
and the query cost increases from 8.66$\mu$s to 15.98$\mu$s
when the triangulation quality is diminished.  
It shows that our query algorithm is less sensitive to triangle quality than BVH. 

\begin{figure}[ht]
    \centering
    \includegraphics[width=0.95\columnwidth]{"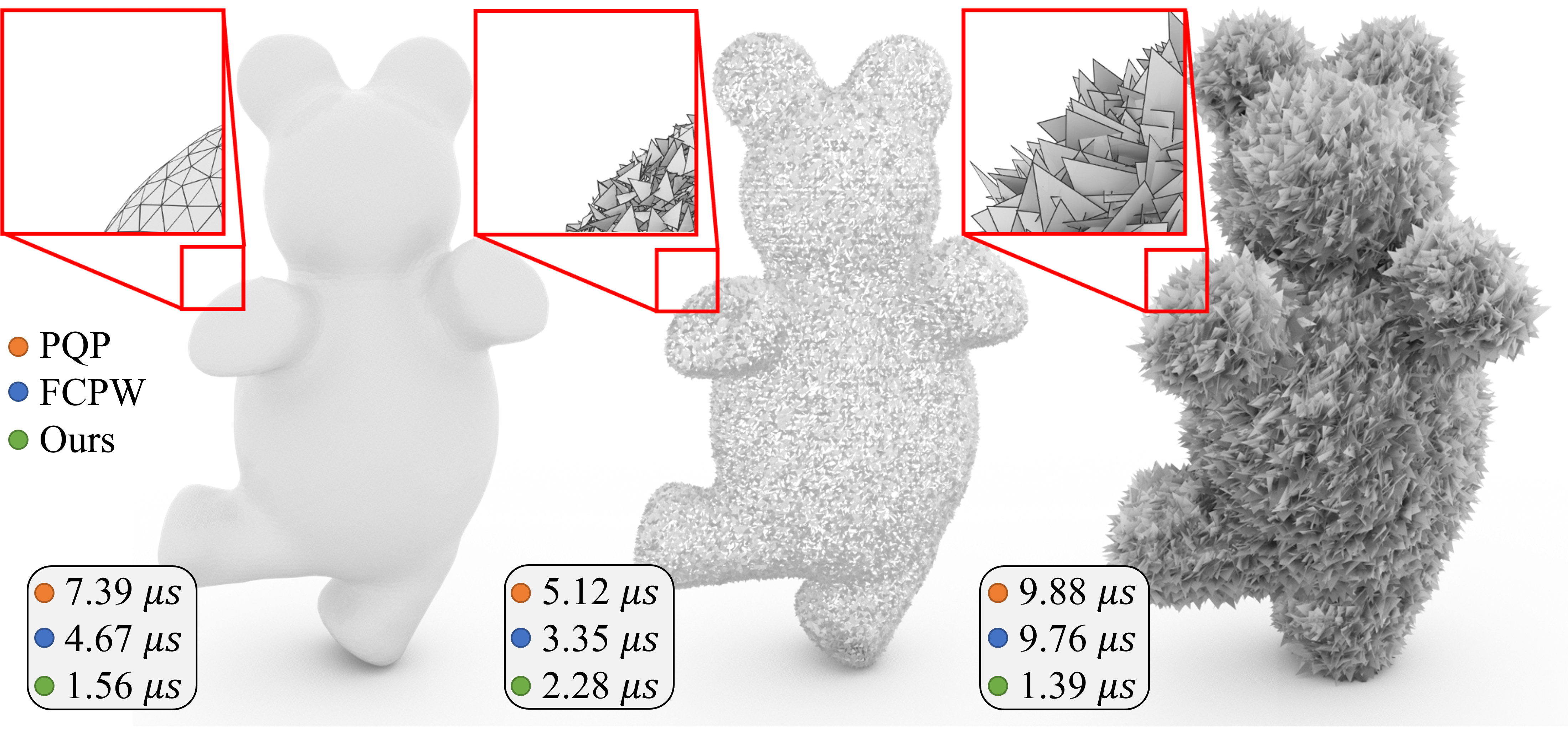"}
    \makebox[0.31\columnwidth][c]{(a)}
    \makebox[0.31\columnwidth][c]{(b)}
    \makebox[0.31\columnwidth][c]{(c)}
    \caption{
    Test PQP, FCPW and ours on the Bear model with watertight/broken triangulation. (a) Watertight triangulation.
    (b) Triangle soup with gapped triangles.
    (c) Triangle soup with high penetration.
    Our comparative advantage over BVH-based methods becomes smaller for a set of gapped triangles whereas larger in the presence of high penetration.
    }
    \label{fig:tri_soup}
    \vspace{-3mm}
\end{figure}
\paragraph{Triangle soup}
Figure~\ref{fig:tri_soup} shows three versions of the 20K-face Bear model, i.e.,
watertight triangle mesh, 
gapped triangles
and a triangle soup with high penetration.
The statistics of query speed are available in Figure~\ref{fig:tri_soup}.
For the watertight Bear model,
our query algorithm runs about 5 times as fast as PQP, and 3 times as fast as FCPW.
Figure~\ref{fig:tri_soup}(b) and Figure~\ref{fig:tri_soup}(c) show that the comparative advantage over them becomes smaller for gapped triangles whereas becomes larger in the presence of high penetration. 

\paragraph{Mixed primitives}
We make tests on a mixed set of segments and triangles.

By combining the triangular mesh of Armadillo model and the wireframe of Camel model,
we synthesize a mixed set of geometric primitives. 
Experimental results show that 
our algorithm is 3.8 times as fast as FCPW \new{(PQP does not support this kind of input)}. 

\subsection{Tests on Thingi10K}
Thingi10K~\cite{thingi10k} is a large scale 3D dataset 
that contains diverse models.

\begin{figure}[ht]
    \centering
    \includegraphics[width=\linewidth]{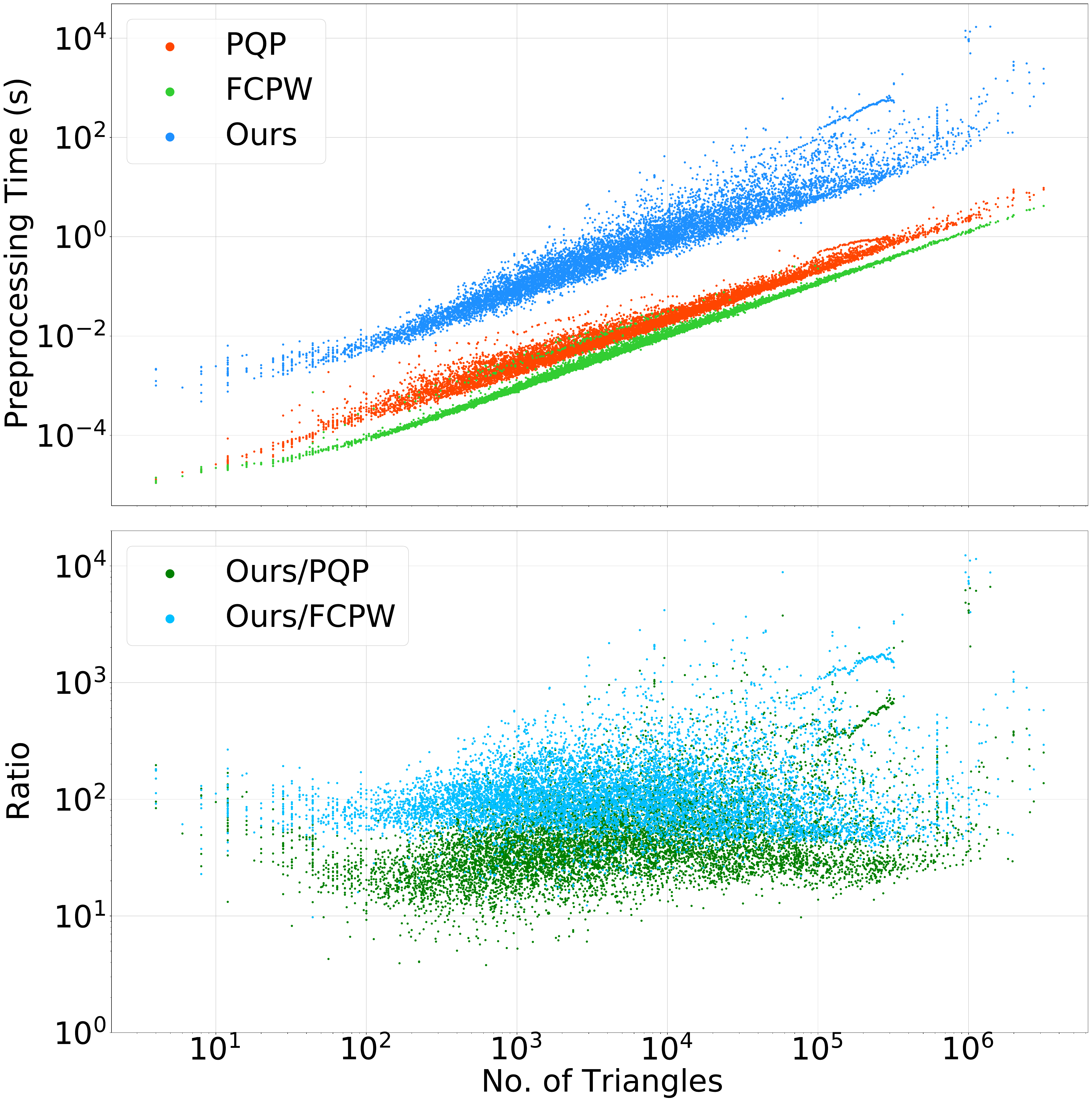}
    \caption{Comparison of preprocessing cost among PQP, FCPW and our algorithm on the full Thingi10K dataset. (log-log plot) }
    \label{fig:Thingi10K_build}
    \vspace{-3mm}
\end{figure}

\begin{figure}[ht]
    \centering
    \includegraphics[width=\linewidth]{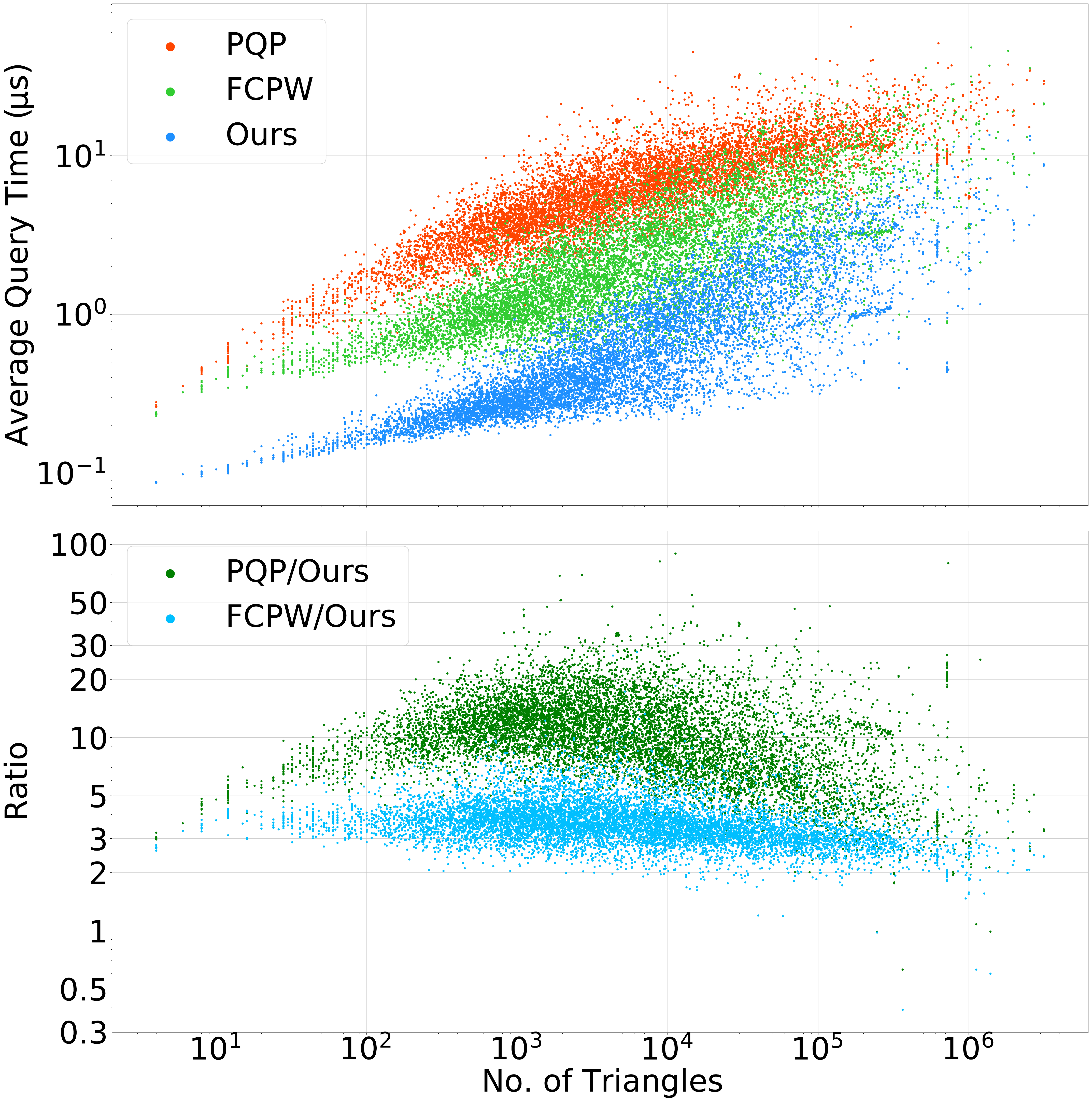}
    \caption{Comparison of query time among PQP, FCPW and our algorithm on the full Thingi10K dataset. (log-log plot) }
    \label{fig:Thingi10K_query}
    \vspace{-3mm}
\end{figure}

\begin{figure}[ht]
    \centering
    \includegraphics[width=\linewidth]{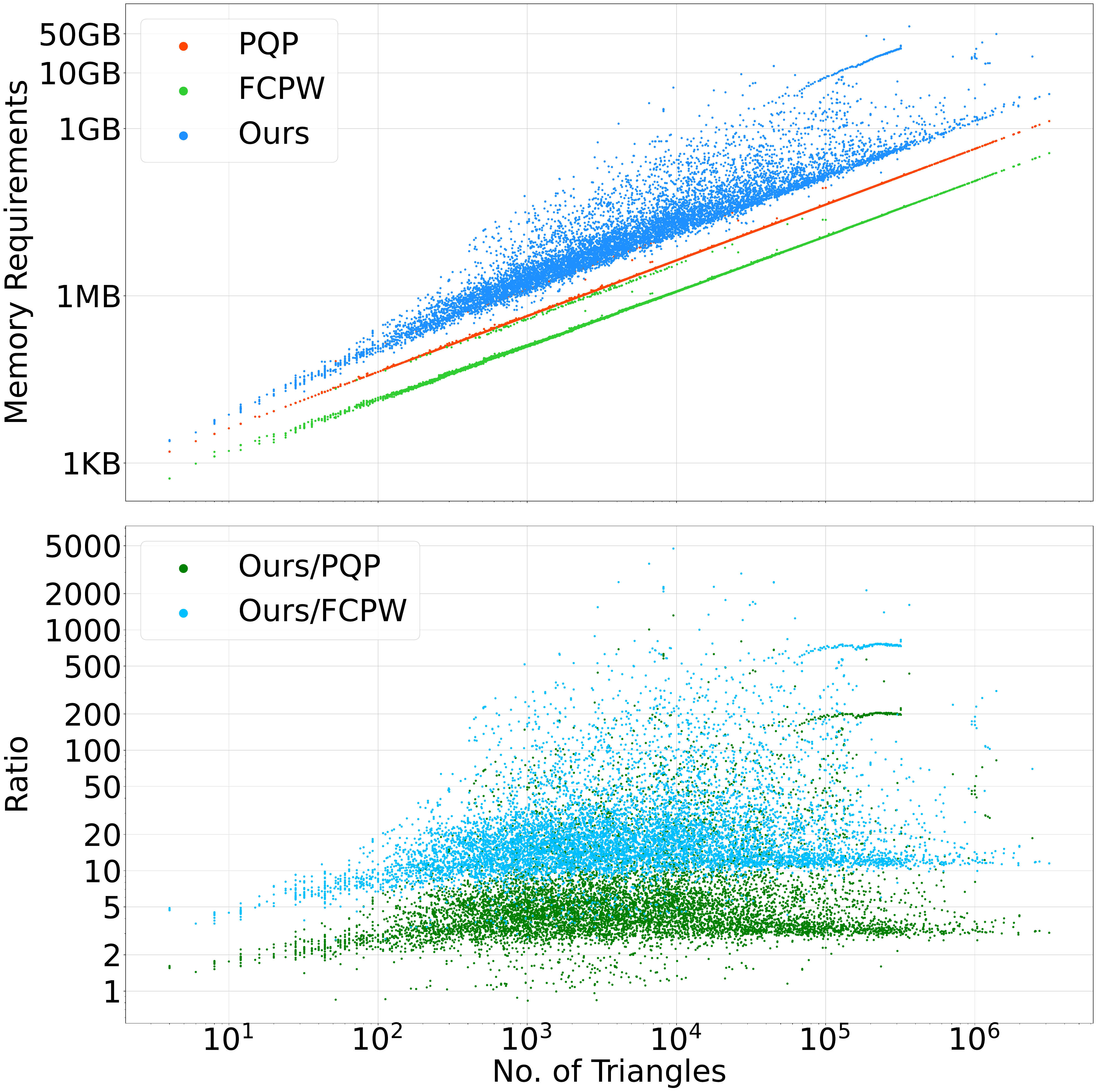}
    \caption{\new{Comparison of memory usage among PQP, FCPW and our algorithm on the full Thingi10K dataset. (log-log plot) }}
    \label{fig:Thingi10K_memory}
    \vspace{-3mm}
\end{figure}

In order to make a comprehensive comparison among PQP, FCPW and our algorithm,
we run them on the full Thingi10K dataset to compare the preprocessing cost (see Figure~\ref{fig:Thingi10K_build}), the query cost (see Figure~\ref{fig:Thingi10K_query}) and the memory requirements (see Figure~\ref{fig:Thingi10K_memory}). 
Based on the statistics, 
we observe that 
our algorithm achieves a higher 
query performance at the cost of preprocessing and memory usage. 
Generally speaking, 
our algorithm runs 2 to 10 times faster than PQP and FCPW.

\section{Conclusion and Limitations}
In this paper, we develop a novel algorithmic paradigm,
named P2M,
to solve the problem of point-to-mesh distance query.
P2M needs to precompute a pair of data structures 
including a KDT of mesh vertices
and an interception table that
encodes the principal-agent relationship between vertices and edges/faces,
such that the query stage proceeds by first searching the KDT and then looking up the interception table to retrieve the closest geometric primitive. 
We give rigorous proofs about the correctness and
propose a set of strategies 
for speeding up the preprocessing stage and the query stage.
We conduct extensive experiments to evaluate our approach. 
Experimental results show that
our algorithm runs many times faster than the SOTAs. 

\new{However, in its current state, our algorithm 
still needs comprehensive improvement.}
First, the construction of the interception table is still time-consuming. 
\new{Statistics show that in the time-consuming interception inspection phase, about 85\% of visited vertices are checked but found not to be an interceptor during flooding. One potential research direction is to quickly exclude the non-interceptor vertices by some filtering techniques.}
Second, the interception table becomes very long 
for a highly symmetrical shape. 
\new{For example, if the input is a spherical surface,
then each vertex intercepts any triangle. 
}
Last but not least, 
PQP supports closest point query, line-surface intersection and collision detection at the same time,
but our algorithm only supports closest point query. 
\new{In the future, we shall further improve the algorithm in terms of the above-mentioned aspects.}

\new{
\section{Acknowledgements}
The authors would like to thank the anonymous reviewers for their valuable comments and suggestions. This work is supported by National Key R\&D Program of China (2021YFB1715900), National Natural Science Foundation of China (62002190, 62272277, 62072284), and Natural Science Foundation of Shandong Province (ZR2020MF036). 
Besides, the authors would like to express their gratitude to Professor Charlie C.L. Wang for providing a modified version of PQP that supports point-to-mesh distance queries.
}

\bibliographystyle{ACM-Reference-Format}
\bibliography{sample-base}

\newpage
\onecolumn
\appendix

\section{Complete Pseudocode}

\begin{algorithm_mine_def}[caption={Preprocess.}, label={alg:preprocess}]
Input: vertices $V$, edges $E$, triangular faces $F$
Output: vertical space $Space^{\perp}$ of $E$ and $F$, KD tree of vertices $KDT$, interception table $IT$, a set of R-trees $RT$
begin
    foreach $f\in F$
        foreach bounding edge $e$ of $f$
            Compute the equation of the vertical plane $\pi$ defined by $e$ and $f$; 
            Record the two half spaces divided by $\pi$ into $Space^{\perp}(e)$ and $Space^{\perp}(f)$ respectively;
        end
    end
    foreach $e\in E$
        Compute the equations of two vertical planes rooted at the endpoints of $e$;
        Record the two half spaces defined by the two planes into $Space^{\perp}(e)$;
    end
    Compute Voronoi diagram of $V$;
    Compute $KDT$ of $V$;
    foreach $e\in E$
        Inspect the vertices in $V$ from $e$'s endpoints in a flooding fashion and add $e$'s interceptors into the interception table $IT$; 
        foreach $e$'s interceptor $v$
            Calculate the bounding box of $\textit{Region}(v,e)$;
        end
    end
    foreach $f\in F$
        Inspect the vertices in $V$ from the three vertices of $f$ in a flooding fashion and add $f$'s interceptors into the interception table $IT$; 
        foreach $f$'s interceptor $v$
            Calculate the bounding box of $\textit{Region}(v,f)$;
        end
    end
    foreach $v\in V$
        Organize the bounding boxes of the primitives intercepted by $v$ into an R-tree;
    end
end
\end{algorithm_mine_def}

\begin{algorithm_mine_def}[caption={Query.}, label={alg:query}]
Input: query point $p$, KD tree of vertices $KDT$, interception table $IT$, corresponding R-trees $RT$
Output: distance $d_{min}$ from $p$ to mesh surface
begin
    Find the closest vertex $v$ to $p$ by $KDT$; 
    Search the R-tree of $v$ and record the intercepted primitives whose corresponding bounding box contains $p$;
    $d_{min}\gets d(p,v)$;
    foreach recorded primitive
        if it is an edge $e$
            if $p$ is inside of the two vertical planes rooted at the endpoints of $e$
                set $d$ as the distance between $p$ and the straight line of $e$
            end
        end
        if it is a face $f$
            if $p\in Space^{\perp}(f)$ 
                set $d$ as the distance between $p$ and the plane of $f$
            end
        end
        $d_{min}\gets MIN(d_{min},d)$
    end
end
\end{algorithm_mine_def}

\end{document}